\documentclass[12pt, english]{article}

\makeatletter

\providecommand{\tabularnewline}{\\}

\@ifundefined{date}{}{\date{}}
\usepackage{inputenc}      % This solve the inputenc error in lyx. If error, use: \usepackage[utf8]{inputenc}
\usepackage[english]{babel}  % solve the error: You haven't loaded the option english yet
\usepackage[letterpaper, margin=1.1in]{geometry}  % margin, paper size etc.
\usepackage{bm}\usepackage{bbm}   % bm for bold math, bbm for \mathbbm{1}
\usepackage{amsmath}   % for \overset, \underset, \text, \lVert, \rVert
\usepackage{amssymb}   % amssymb or amsfonts for \mathbb
\usepackage{graphicx}  % \includegraphics, \resizebox. graphicx is better than graphics.
\usepackage{cases}     % \begin{cases} for stack of equations after the bracket
\usepackage{multirow}  % \multirow and \multicolumn
\usepackage{longtable} % \begin{longtable} for long tables.
\usepackage{caption}   % \caption*{}, \captionof (while using \resize)
\usepackage{float}     % prevent table from floating by \begin{table}[H]. Using [h] without this package does not avoid the floating.
\usepackage{adjustbox} % \begin{adjustbox} to resize the table
\usepackage[title]{appendix} % more control of appendix
\usepackage{array}     % vertical center in the table \begin{tabular}{ | m{1in} |... }
\usepackage{xcolor}    % change text color
\usepackage{url}       % include url: \url{}

\newcommand{\vone}{\mathbbm{1}}      % command to write vector 1 in vector form
\newcommand{\norm}[1]{\left\| #1 \right\|}   % command to write norm
\newcommand{\w}{\widetilde}          % command to write wide tilde

\usepackage{enumitem}  % Change label: \begin{enumerate}[label=(\arabic*)] . Not compatible with lyx: Remove [label=(\arabic*)].
\usepackage{hhline}    % \hhline double lines in tables. Not compatible with lyx: Replace \\hhline\{.*\}\} with \\hline, replace \\hline\{.*\} with \\hline.

\begin{document}

\title{\vspace{-0.5in}
High-Dimensional Estimation, Basis Assets, and the Adaptive Multi-Factor Model
\thanks{We thank the editors, referees, Dr. Manny Dong, and all the support
from Cornell University.}}

\author{Liao Zhu\thanks{Ph.D. in the Department of Statistics and Data Science, Cornell University, Ithaca, New York 14853, USA. % Email: lz384@cornell.edu. Phone: 607-379-7330.
},
Sumanta Basu\thanks{Shayegani Bruno Family Faculty Fellow and Assistant Professor of Department of Statistics and Data Science, Cornell University, Ithaca, New York 14853, USA. % Email: sumbose@cornell.edu. Phone: 734-358-7953.
},
Robert A. Jarrow\thanks{Corresponding author: Robert Jarrow, Ronald P. and Susan E. Lynch Professor of Investment Management, Samuel Curtis Johnson Graduate School of Management, Cornell University, Ithaca, New York 14853, USA and Kamakura Corporation, Honolulu, Hawaii 96815. Email: raj15@cornell.edu. % Phone: 607-539-6070.
},
Martin T. Wells\thanks{Charles A. Alexander Professor of Statistical Sciences, Department
of Statistics and Data Science, Cornell University, Ithaca, New York 14853, USA. % Email: mtw1@cornell.edu. Phone: 607-255-8801.
} }

\date{}

\maketitle
\global\long\def\thefootnote{\arabic{footnote}}
%\blindtext

%\tableofcontents
\begin{abstract}
The paper proposes a new algorithm for the high-dimensional financial data -- the Groupwise Interpretable Basis Selection (GIBS) algorithm, to estimate a new Adaptive Multi-Factor (AMF) asset pricing model, implied by the recently developed Generalized Arbitrage Pricing Theory, which relaxes the convention that the number of risk-factors is small. We first obtain an adaptive collection of basis assets and then simultaneously test which basis assets correspond to which securities, using high-dimensional methods. The AMF model, along with the GIBS algorithm, is shown to have a significantly better fitting and prediction power than the Fama-French 5-factor model.
\end{abstract}
%\includepdf{Disclosure_Liao_Zhu_signed}
%\includepdf{Disclosure_Sumanta_Basu_signed}
%\includepdf{Disclosure_Robert_A_Jarrow_signed}
%\includepdf{Disclosure_Martin_T_Wells_signed}
\clearpage

\section{Introduction}

The purpose of this paper is to proposes a new algorithm for the high-dimensional \\financial data, the \textbf{Groupwise Interpretable Basis Selection (GIBS) algorithm} to estimate a new \textbf{Adaptive Multi-Factor (AMF) asset pricing model}, implied by the generalized arbitrage pricing theory (APT) recently developed by Jarrow and Protter (2016) \cite{jarrow2016positive} and Jarrow (2016) \cite{jarrow2016bubbles}. \footnote{The Appendix \ref{ap: summary_apt} provides a brief summary of the generalized APT.} This generalized APT is derived in a continuous time, continuous trading economy imposing only the assumptions of frictionless markets, competitive markets, and the existence of a martingale measure.\footnote{By results contained in Jarrow and Larsson (2012) \cite{jarrow2012meaning}, this is equivalent to the economy satisfying no free lunch with vanishing risk (NFLVR) and no dominance (ND). See Jarrow and Larrson (2012) \cite{jarrow2012meaning} for the relevant definitions.} As such, this generalized APT includes both Ross's (1976) \cite{ross1976arbitrage} static APT and Merton's (1973) \cite{merton1973intertemporal} inter-temporal CAPM as special cases.

The generalized APT has four advantages over the traditional APT and
the inter-temporal CAPM. First, it derives the \emph{same} form of
the empirical estimation equation (see Equation (\ref{eq: amf_theory_ols_subset})
below) using a weaker set of assumptions, which are more likely to
be satisfied in practice.\footnote{The stronger assumptions in Ross's APT are: (i) a realized return
process consisting of a finite set of common factors and an idiosyncratic
risk term across a countably infinite collection of assets, and (ii)
no infinite asset portfolio arbitrage opportunities; in Merton's ICAPM
they are assumptions on (i) preferences, (ii) endowments, (ii) beliefs
and information, and (iv) those necessary to guarantee the existence
of a competitive equilibrium. None of these stronger assumptions are
needed in Jarrow and Protter (2016) \cite{jarrow2016positive}.} Second, the no-arbitrage relation is derived with respect to realized
returns, and not with respect to expected returns. This implies, of
course, that the error structure in the estimated multi-factor model
is more likely to lead to a larger $R^{2}$ and to satisfy the standard
assumptions required for regression models. Third, the set of basis
assets and the implied risk-factors are tradeable under the generalized
APT, implying their potential observability. Fourth, since the space
of random variables generated by the uncertainty in the economy is
infinite dimensional, the implied basis asset representation of any
security's return is parsimonious and sparse. Indeed, although the
set of basis assets is quite large (possibly infinite dimensional),
only a finite number of basis assets are needed to explain any assets'
realized return and different basis assets apply to different assets.
This last insight is certainly consistent with intuition since an
Asian company is probably subject to different risks than is a U.S.
company. Finally, adding a non-zero alpha to the no-arbitrage relation
in realized return space enables the identification of arbitrage opportunities.
This last property is also satisfied by the traditional APT and the
inter-temporal CAPM.

The generalized APT is important for practice because it provides
an exact identification of the relevant set of basis assets characterizing
a security's \emph{realized} (emphasis added) returns. This enables
a more accurate risk-return decomposition facilitating its use in
trading (identifying mispriced assets) and for risk management. Taking
expectations of this realized return relation with respect to the
martingale measure determines which basis assets are \emph{risk-factors},
i.e. which basis assets have non-zero expected excess returns (risk
premiums) and represent systematic risk. Since the traditional models
are nested within the generalized APT, an empirical test of the generalized
APT provides an alternative method for testing the traditional models
as well. One of the most famous empirical representations of a multi-factor
model is given by the Fama-French (2015) \cite{fama2015five}
five-factor model (FF5), see also \cite{fama1993common, fama1992cross}.
Recently, Harvey, Liu, and Zhu (2016) \cite{harvey2016and}
reviewed the literature on the estimation of factor models, the collection
of risk-factors employed, and argued for the need to use an alternative
statistical methodology to sequentially test for new risk-factors.
Our paper provides one such alternative methodology using the collection
of basis assets to determine which of these earn risk premium.

Since the generalized APT is a model for \textit{realized} returns
that allows different basis assets to affect different stocks differently,
an empirical test of this model starts with slightly different goals
than tests of conventional asset pricing models (discussed above)
whose implications are only with respect to \textit{expected} returns
and risk-factors. First, instead of searching for a few common risk-factors
that affect the entire cross-section of expected returns, as in the
conventional approach, we aim to find an exhaustive set of basis assets,
while maintaining parsimony for each individual stock (and hopefully
for the cross-section of stocks as well), using the GIBS algorithm we propose here. This alternative approach
has the benefit of increasing the explained variation in our time
series regressions. Second, as a direct implication of the estimated
realized return relation, the cross-section of expected returns is
uniquely determined. This implies, of course, that the collection
of risk-factors will be those basis assets with non-zero expected
excess returns (i.e. they earn non-zero risk premium).\footnote{An investigation of the cross-section of expected returns implied
by the generalized APT model estimated in this paper is a fruitful
area for future research. }

In addition, compared to PCA based methods that construct risk-factors
from linear combination of \textit{various} stocks, which are consequently
often difficult to interpret, our GIBS algorithm consists of a set of \textit{interpretable
and tradeable} basis assets. The new model explains more variation
in \textit{realized} returns. It is important to note here that in
a model of realized returns, some of the basis assets will reflect
idiosyncratic risks that do not earn non-zero risk premium. Those
basis assets that have non-zero excess expected returns are the relevant
risk-factors identified in the traditional estimation methodologies.
While a few recent papers adopt high-dimensional estimation methods
for modeling the cross-section of expected returns and an associated
parsimonious representation of the stochastic discount function \cite{bryzgalova2015spurious,feng2017taming},
our empirical test is specifically designed to align with the generalized
APT model's implications using basis assets and realized returns.

To test the generalized APT, we first obtain the collection of all
possible basis assets. Then, we provide a simultaneous test, security
by security, of which basis assets are significant for each security.
However, there are several challenges that must be overcome to execute
this estimation. First, in the security return regression using the
basis assets as independent variables, due to the assumption that
the regression coefficients ($\beta$'s) are constant, it's necessary
to run the estimation over a small time window because the $\beta$'s
are likely to change over longer time windows. This implies that the
number of sample points may be less than the number of independent
variables ($p>n$). This is the so-called high-dimension regime, where
the ordinary least squares (OLS) solution no longer holds. Second,
the collection of basis assets selected for investigation will be
highly correlated. And, it is well known that large correlation among
independent variables causes difficulties (redundant basis assets
selected, low fitting accuracy etc., see \cite{hebiri2012how,vandegeer2013lasso})
in applying the Least Absolute Shrinkage and Selection Operator (LASSO).
To address these difficulties, we propose a novel and hybrid algorithm
-- the GIBS algorithm, for identifying basis assets which are different
from the traditional variance-decomposition approach. The GIBS algorithm
takes advantage of several high-dimensional methodologies, including
prototype clustering, LASSO, and the ``1se rule'' for prediction.

We investigate Exchange-Traded Funds (ETFs) since they inherit and
aggregate the basis assets from their constituents. In recent years
there are more than one thousand ETFs, so it is reasonable to believe
that one can obtain the basis assets from the collection of ETFs in
the CRSP database, plus the Fama-French 5 factors. Consider the market
return, one of the Fama-French 5 factors. It can be duplicated by
ETFs such as the SPDR S\&P 500 ETF, the Vanguard S\&P 500 ETF, etc.
Therefore, it is also reasonable to believe that the remaining ETFs
can represent other basis assets as well.

We group the ETFs into different asset classes and use prototype clustering
to find good representatives within each class that have low pairwise
correlations. This reduced set of ETFs forms our potential basis assets.
After finding this set of basis assets, we still have more basis assets
than observations ($p>n$), but the basis assets are no longer highly
correlated. This makes LASSO an appropriate approach to determine
which set of basis assets are important for a security's return. To
be consistent with the literature, we fit an OLS regression on each
security's return with respect to its basis assets (that are selected
by LASSO) to perform an intercept ($\alpha$) and a goodness of fit
test. The importance of these tests are discussed next.

As noted above, the intercept test can be interpreted as a test of
the generalized APT under the assumptions of frictionless, competitive,
and arbitrage-free markets (more formally, the existence of an equivalent
martingale measure). The generalized APT abstracts from market microstructure
frictions, such as bid-ask spreads and execution speeds (costs), and
strategic trading considerations, such as high-frequency trading.
To be consistent with this abstraction, we study returns over a weekly
time interval, where the market microstructure frictions and strategic
trading considerations are arguably less relevant. Because the generalized
APT ignores market microstructure considerations, we label it a ``large-time
scale'' model.

If we fail to reject a zero alpha, we accept this abstraction, thereby
providing support for the assertion that the frictionless, competitive,
and arbitrage-free market construct is a good representation for ``large
time scale'' security returns. If the model is accepted, a goodness
of fit test quantifies the explanatory power of the model relative
to the actual time series variations in security returns. A ``good''
model is one where the model error (the difference between the model's
predictions and actual returns) behaves like white noise with a ``small''
variance. The adjusted $R^{2}$ provides a good metric of comparison
in this regard. Conversely, if we reject a zero alpha, then this is
evidence consistent with either: (i) that microstructure considerations
are necessary to understand ``large time scale'' models, or (ii)
that there exist arbitrage opportunities in the market. This second
possibility is consistent with the generalized APT being a valid description
of reality, but where markets are inefficient. To distinguish between
these two alternatives, we note that a non-zero intercept enables
the identification of these ``alleged'' arbitrage opportunities,
constructed by forming trading strategies to exploit the existence
of these ``positive alphas''. The implementation of these trading
strategies enables a test between these two alternatives.

Here is a brief summary of our results.
\begin{itemize}
\item The AMF model gives fewer significant intercepts (alphas) as compared to the Fama-French 5-factor model (percentage of companies with non-zero intercepts from 6.33\% to 4.12\% ). For both models, considering the False Discovery Rate, we cannot reject the hypothesis that the intercept is zero for all securities in the sample. This implies that historical security returns are consistent with the behavior implied by ``large-time scale'' models.
\item In an Goodness-of-Fit test comparing the Fama-French 5-factor and the AMF model, the AMF model has a substantially larger In-Sample Adjusted $R^{2}$ and the difference of goodness-of-fit of two models are significant. Furthermore, the AMF model increased the Out-of-Sample $R^2$ for the prediction by $ 58\% $. This supports the superior performance of the generalized APT in characterizing security returns.
\item As a robustness test, for those securities whose intercepts were non-zero (although insignificant), we tested the AMF model to see if positive alpha trading strategies generate arbitrage opportunities. They do not, thereby confirming the validity of the generalized APT.
\item The estimated GIBS algorithm selects 186 basis assets for the AMF model. All of these basis assets are significant for some stock, implying that a large number of basis assets are needed to explain security returns. On average each stock is related to only 3.4 basis assets, with most stocks having between $1\sim10$ significant basis assets. Cross-validation results in the Section \ref{sec: amf_comparison} are consistent with our sparsity assumption. Furthermore, different securities are related to different basis assets, which can be seen in Table \ref{tab: amf_perc_sgn} and the Heat Map in Figure \ref{fig: amf_heatmap}. Again, these observations support the validity of the generalized APT.
\item To identify which of the basis assets are risk-factors, we compute the average excess returns on the relevant basis assets over the sample period. These show that 78\% of the basis assets are risk-factors, earning significant risk premium.
\item Comparison of GIBS with the alternative methods discussed in Section \ref{sec: amf_comparison} shows the superior performance of GIBS. The comparison between GIBS and GIBS + FF5 shows that some of the FF5 factors are overfitting noise in the data.
\end{itemize}
\medskip{}

More recently, insightful papers by Kozak, Nagel, and Santosh (2017,
2018) \cite{kozak2018interpreting, kozak2017shrinking} proposed alternative
methods for analyzing risk-factors models. As Kozak et al. (2018)
\cite{kozak2018interpreting} note, if the ``risk-factors'' are
considered as a variance decomposition for a large amount of stocks,
one can always find that the number of important principal components
is small. However, this may not imply that there are only a small
number of relevant risk-factors because the Principal Component Analysis
(PCA) can either mix the underlying risk-factors together or separate
them into several principal components. It may be that there are a
large number of risk-factors, but the ensemble appears in only a few
principal components. The sparse PCA method used in \cite{kozak2018interpreting}
removes many of the weaknesses of traditional PCA, and even gives an interpretation of the risk-factors as stochastic discount risk-factors. However all these methods still suffer from the problems (low interpretability, low prediction accuracy, etc.) inherited from the variance decomposition framework. An alternative approach, the one we use here, is to abandon variance decomposition methods and to use high-dimensional methods instead,
such as prototype clustering to select basis assets as the ``prototypes'' or ``center'' of the groups they are representing. The proposed GIBS method gives much clearer interpretation, and much better prediction accuracy.

A detailed comparison of alternative methods that address the difficulties
of high-dimension and strong correlation id given in Section \ref{sec: amf_comparison}
below. Other methods include Elastic Net or Ridge Regression to deal
with the correlation. Elastic Net and Ridge Regression handle multicollinearity
by adding penalties to make the relevant matrix invertible. However,
neither of these methods considers the underlying cluster structure,
which makes it hard to interpret the selected basis assets. This is
the reason for the necessity of prototype clustering in the first
step, since it gives an interpretation of the selected basis assets
as the ``prototypes'' or the ``centers'' of the groups they are
representing. After the prototype clustering, we use a modified version
of LASSO (use in the GIBS algorithm) instead of either Elastic Nets
or Ridge Regression. The reason is that, compared to GIBS, alternative
methods achieve a much lower prediction goodness-of-fit (see Table
\ref{tab: amf_compare_methods}), but select more basis assets (see Figure
\ref{fig: amf_compare_n_factors}), which overfits and makes the model
less interpretable.

The tuning parameter that controls sparsity in LASSO, $\lambda$,
is traditionally selected by cross-validation and with this $\lambda$,
the model selects an average of 13.4 basis assets for each company.
However, as shown in the comparison Section \ref{sec: amf_comparison},
this overfits the noise in the data when compared with the GIBS algorithm
with respect to Out-of-Sample $R^{2}$. The reasons for the poor performance
of this cross-validation is discussed in Section \ref{sec: amf_comparison}
below. Consequently, to control against overfitting, we use the ``1se
rule'' along with the threshold that the number of basis assets can
not exceed 20.

The Adaptive Multi-Factor (AMF) model estimated by the GIBS algorithm
in this paper is shown to be consistent with the data and superior
to the Fama-French 5-factor model. An outline of this paper is as
follows. Section \ref{sec: amf_high_dim_method} describes the high-dimensional
statistical methods used in this paper and Section \ref{sec: amf_model}
presents the Adaptive Multi-Factor (AMF) model to be estimated. Section
\ref{sec: gibs_algo} gives the proposed GIBS algorithm to estimate
the model and Section \ref{sec: amf_results} presents the empirical
results. Section \ref{sec: amf_comparison} discusses the reason we
chose our method and provide a detailed comparison over alternative
methods. Section \ref{sec: amf_risk_premium} discusses the risk premium
of basis assets and Section \ref{sec: amf_illustration} presents some
illustrative examples. Section \ref{sec: amf_conclusion} concludes. All codes are written in R and can be found on Github \url{https://github.com/pig618/amf_gibs}.

\section{High-Dimensional Statistical Methodology}
\label{sec: amf_high_dim_method}

Since high-dimensional statistics is relatively new to the finance
literature, this section reviews the relevant statistical methodology.

\subsection{Preliminaries and Notations}
\label{sec: notations}

Let $\norm{\bm{v}}_{q}$ denote the standard $l_{q}$ norm of a vector
$\bm{v}$ of dimension $p\times1$, i.e.
\[
\norm{\bm{v}}_{q}=\begin{cases}
(\sum_{i}|\bm{v}_{i}|^{q})^{1/q} & \text{if}\;\;0<q<\infty\\
\#\{i:\bm{v}_{i}\neq0\}, & \text{if}\;\;q=0\\
\max_{i}|\bm{v}_{i}| & \text{if}\;\;q=\infty.
\end{cases}
\]
Suppose $\bm{\beta}$ is also a vector with dimension $p\times1$,
a set $S\subseteq\{1,2,...,p\}$, then $\beta_{S}$ is a $p\times1$
vector with i-th element
\[
(\bm{\beta}_{S})_{i}=\begin{cases}
\beta_{i}, & \text{if}\;\;i\in S\\
0, & otherwise.
\end{cases}
\]

Here the index set $S$ is called the support of $\bm{\beta}$, in
other words, $supp(\bm{\beta})=\{i:\bm{\beta}_{i}\neq0\}$. Similarly,
if $\bm{X}_{n\times m} $ is a matrix instead of a vector, for any index set $ S \subseteq\{1,2,..., m\} $, use $\bm{X}_{S}$ to denote the the columns of $\bm{X}$ indexed by $S$. Denote $\vone_{n}$ as a $n\times1$ vector with all elements being 1, $\bm{J_{n}}=\vone_{n}\vone_{n}'$ and $\bm{\bar{J}_{n}}=\frac{1}{n}\bm{J_{n}}$. $\bm{I_{n}}$ denotes the identity matrix with diagonal 1 and 0 elsewhere. The subscript
$n$ is always omitted when the dimension $n$ is clear from the context. The notation $\#S$ means the number of elements in the set $S$.

\subsection{Minimax Prototype Clustering and Lasso Regressions}
\label{sec: proto_lasso}

This section describes the prototype clustering to be used to deal
with the problem of high correlation among the independent variables
in our LASSO regressions. To remove unnecessary independent variables,
using clustering methods, we classify them into similar groups and
then choose representatives from each group with small pairwise correlations.
First, we define a distance metric to measure the similarity between
points (in our case, the returns of the independent variables). Here,
the distance metric is related to the correlation of the two points,
i.e.
\begin{equation}
d(r_{1},r_{2})=1-|corr(r_{1},r_{2})|\label{eq: corr_dist}
\end{equation}
where $r_{i}=(r_{i,t},r_{i,t+1},...,r_{i,T})'$ is the time series
vector for independent variable $i=1,2$ and $corr(r_{1},r_{2})$
is their correlation. Second, the distance between two clusters needs
to be defined. Once a cluster distance is defined, hierarchical clustering
methods (see \cite{kaufman2009finding}) can be used to organize the
data into trees.

In these trees, each leaf corresponds to one of the original data
points. \\ Agglomerative hierarchical clustering algorithms build trees
in a bottom-up \\ approach, initializing each cluster as a single point,
then merging the two closest clusters at each successive stage. This
merging is repeated until only one cluster remains. Traditionally,
the distance between two clusters is defined as either a complete
distance, single distance, average distance, or centroid distance.
However, all of these approaches suffer from interpretation difficulties
and inversions (which means parent nodes can sometimes have a lower
distance than their children), see Bien, Tibshirani (2011)\cite{bien2011hierarchical}.
To avoid these difficulties, Bien, Tibshirani (2011)\cite{bien2011hierarchical}
introduced hierarchical clustering with prototypes via a minimax linkage
measure, defined as follows. For any point $x$ and cluster $C$,
let
\begin{equation}
d_{max}(x,C)=\max_{x'\in C}d(x,x')
\end{equation}
be the distance to the farthest point in $C$ from $x$. Define the
\emph{minimax radius} of the cluster $C$ as
\begin{equation}
r(C)=\min_{x\in C}d_{max}(x,C)
\end{equation}
that is, this measures the distance from the farthest point $x\in C$
which is as close as possible to all the other elements in C. We call
the minimizing point the \emph{prototype} for $C$. Intuitively, it
is the point at the center of this cluster. The \emph{minimax linkage}
between two clusters $G$ and $H$ is then defined as
\begin{equation}
d(G,H)=r(G\cup H).
\end{equation}
Using this approach, we can easily find a good representative for
each cluster, which is the prototype defined above. It is important
to note that minimax linkage trees do not have inversions. Also, in
our application as described below, to guarantee interpretable and
tractability, using a single representative independent variable is
better than using other approaches (for example, principal components
analysis (PCA)) which employ linear combinations of the independent
variables.

The LASSO method was introduced by Tibshirani (1996) \cite{tibshirani1996regression}
for model selection when the number of independent variables ($p$)
is larger than the number of sample observations ($n$). The method
is based on the idea that instead of minimizing the squared loss to
derive the Ordinary Least Squares (OLS) solution for a regression, we should add to the loss
a penalty on the absolute value of the coefficients to minimize the
absolute value of the non-zero coefficients selected. To illustrate
the procedure, suppose that we have a linear model
\begin{equation}
\bm{y}=\bm{X\beta}+\bm{\epsilon}\qquad\text{where}\qquad\bm{\epsilon}\sim N(0,\sigma_{\epsilon}^{2}\bm{I}),\label{eq: lasso}
\end{equation}
$\bm{X}$ is an $n\times p$ matrix, $\bm{y}$ and $\bm{\epsilon}$
are $n\times1$ vectors, and $\bm{\beta}$ is a $p\times1$ vector.

The LASSO estimator of $\bm{\beta}$ is given by
\begin{equation}
\bm{\hat{\beta}}_{\lambda}=\underset{\bm{\beta}\in\mathbb{R}^{p}}{\arg\min}\left\{ \frac{1}{2n}\norm{\bm{y}-\bm{X\beta}}_{2}^{2}+\lambda\norm{\bm{\beta}}_{1}\right\} \label{eq: lasso_beta}
\end{equation}
where $\lambda>0$ is the tuning parameter, which determines the magnitude
of the penalty on the absolute value of non-zero $\beta$'s. In this
paper, we use the R package \textit{glmnet} \cite{friedman2010regularization}
to fit LASSO.

In the subsequent estimation, we will only use a modified version
of LASSO as a model selection method to find the collection of important
independent variables. After the relevant basis assets are selected,
we use a standard Ordinary Least Squares (OLS) regression on these variables to test for the
goodness of fit and significance of the coefficients. More discussion
of this approach can be found in Zhao, Shojaie, Witten (2017) \cite{zhao2017defense}.

In this paper, we fit the prototype clustering followed by a LASSO
on the prototype basis assets selected. The theoretical justification
for this approach can be found in \cite{reid2018general} and \cite{zhao2017defense}.

\section{The Adaptive Multi-Factor Model}

\label{sec: amf_model}

This section presents the Adaptive Multi-Factor (AMF) asset pricing model that is
estimated using the high-dimensional statistical methods just discussed.
Given is a frictionless, competitive, and arbitrage free market. In
this setting, a dynamic generalization of Ross's (1976) \cite{ross1976arbitrage}
APT and Merton's (1973) \cite{merton1973intertemporal} ICAPM derived by Jarrow
and Protter (2016) \cite{jarrow2016positive} implies that the following
relation holds for any security's \emph{realized }return:
\begin{equation}
R_{i}(t)-r_{0}(t)=\sum_{j=1}^{p}\beta_{i,j}\big[r_{j}(t)-r_{0}(t)\big]
=\bm{\beta}_{i}' [\bm{r}(t)-r_{0}(t)\vone]
\label{eq: amf_linear model}
\end{equation}
where at time $t$, $R_{i}(t)$ denotes the return of the \textit{i-th} security
for $1\leq i\leq N$ (where $N$ is the number of securities), $r_{j}(t)$
denotes the return used as the \textit{j-th} basis asset for $1\leq j\leq p$,
$r_{0}(t)$ is the risk free rate, $\bm{r}(t)=(r_{1}(t),r_{2}(t),...,r_{p}(t))'$
denotes the vector of security returns, $\vone$ is a column vector
with every element equal to one, and $\bm{\beta}_{i}=(\beta_{i,1},\beta_{i,2},...,\beta_{i,p})'$.

This generalized APT requires that the basis assets are represented
by traded assets. In Jarrow and Protter (2016) \cite{jarrow2016positive}
the collection of basis assets form an algebraic basis that spans
the set of security payoffs at the model's horizon, time $T$. No
arbitrage, i.e. the existence of a martingale measure, implies that
this same basis set applies to the returns over intermediate time
periods $t\in[0,T]$, which yields the basis asset risk-return relation
given in Equation (\ref{eq: amf_linear model}). It is important to
emphasize that this no-arbitrage relation is for realized returns,
not expected returns. Realized returns are the objects to which asset
pricing estimation is applied. Secondly, the no-arbitrage relation
requires the additional assumptions of frictionless and competitive
markets. Consequently, this asset pricing model abstracts from market
micro-structure considerations. For this reason, this model structure
is constructed to understand security returns over larger time intervals
(days or weeks) and not intra-day time intervals where market micro-structure
considerations apply.

Consistent with this formulation, we use traded ETFs for the basis
assets. In addition, to apply the LASSO method, for each security
$i$ we assume that only a small number of the $\beta_{i,j}$ coefficients
are non-zero ($\beta_{i}$ has the sparsity property). Lastly, to
facilitate estimation, we also assume that the $\beta_{i,j}$ coefficients
are constant over time, i.e. $\beta_{i,j}(t)=\beta_{i,j}$. This assumption
is an added restriction, not implied by the theory. It is only a reasonable
approximation if the time period used in our estimation is not too
long (we will return to this issue subsequently).

To empirically test our model, both an intercept $\alpha_{i}$ and
a noise term $\epsilon_{i}(t)$ are added to Equation (\ref{eq: amf_linear model}),
that is,
\begin{equation}
R_{i}(t)-r_{0}(t)=\alpha_{i}+\sum_{j=1}^{p}\beta_{i,j}(t)\big[r_{j}(t)-r_{0}(t)\big]
+\epsilon_{i}(t)=\alpha+\bm{\beta}_{i}'[\bm{r}(t)-r_{0}(t)\vone]+\epsilon_{i}(t)
\label{eq: amf_testable}
\end{equation}
where $\epsilon_{i}(t)\overset{iid}{\sim}N(0,\sigma_{i}^{2})$ and
$1\leq i\leq N$.

The error term is included to account for noise in the data and ``random''
model error, i.e. model error that is unbiased and inexplicable according
to the independent variables included within the theory. If our theory
is useful in explaining security returns, this error should be small
and the adjusted $R^{2}$ large. The $\alpha$ intercept is called
\emph{Jensen's alpha}. Using the recent theoretical insights of Jarrow
and Protter (2016) \cite{jarrow2016positive}, the intercept test
can be interpreted as a test of the generalized APT under the assumptions
of frictionless, competitive, and arbitrage-free markets (more formally,
the existence of an equivalent martingale measure). As noted above,
this approach abstracts from market microstructure frictions, such
as bid-ask spreads and execution speeds (costs), and strategic trading
considerations, such as high-frequency trading. To be consistent with
this abstraction, we study returns over a weekly time interval, where
the market microstructure frictions and strategic trading considerations
are arguably less relevant. If we fail to reject a zero alpha, we
accept this abstraction, thereby providing support for the assertion
that the frictionless, competitive, and arbitrage-free market construct
is a good representation of ``large time scale'' security returns.
If the model is accepted, a goodness of fit test quantifies the explanatory
power of the model relative to the actual time series variations in
security returns. The adjusted $R^{2}$ provides a good test in this
regard. The GRS test in \cite{gibbons1989test} is usually an excellent
procedure for testing intercepts but it is not appropriate in the
LASSO regression setting.

Conversely, if we reject a zero alpha, then this is evidence consistent
with either: (i) that microstructure considerations are necessary
to understand ``large time scale'' as well as ``short time scale''
returns or (ii) that there exist arbitrage opportunities in the market.
This second possibility is consistent with the generalized APT being
a valid description of reality, but where markets are inefficient.
To distinguish between these two alternatives, we note that a non-zero
intercept enables the identification of these ``alleged'' arbitrage
opportunities, constructed by forming trading strategies to exploit
the existence of these ``positive alphas.''

Using weekly returns over a short time period necessitates the use
of high-dimensional statistics. To understand why, consider the following.
For a given time period $(t,T)$, letting $n=T-t+1$, we can rewrite
Equation (\ref{eq: amf_testable}) using time series vectors as
\begin{equation}
\bm{R}_{i}-\bm{r}_{0}=\alpha_{i}\vone_{n}+(\bm{r}
-\bm{r}_{0}\vone'_{p})\bm{\beta}_{i}+\bm{\epsilon}_{i}
\label{eq: amf_vec}
\end{equation}
where $ 1 \leq i \leq N $, $\bm{\epsilon}_{i}\sim N(0,\sigma_{i}^{2}\bm{I}_{n})$ and {\small
\begin{equation}
\bm{R}_{i}=\begin{pmatrix}R_{i}(t)\\
R_{i}(t+1)\\
\vdots\\
R_{i}(T)
\end{pmatrix}_{n\times1},\,
\bm{r}_{0}=\begin{pmatrix}r_{0}(t)\\
r_{0}(t+1)\\
\vdots\\
r_{0}(T)
\end{pmatrix}_{n\times1},\,
\bm{\epsilon}_{i}=\begin{pmatrix}\epsilon_{i}(t)\\
\epsilon_{i}(t+1)\\
\vdots\\
\epsilon_{i}(T)
\end{pmatrix}_{n\times1}
\end{equation}

\begin{equation}
\bm{\beta}_{i}=\begin{pmatrix}\beta_{i,1}\\
\beta_{i,2}\\
\vdots\\
\beta_{i,p}
\end{pmatrix}_{p\times1}, \,
\bm{r}_{i}=\begin{pmatrix}r_{i}(t)\\
r_{i}(t+1)\\
\vdots\\
r_{i}(T)
\end{pmatrix}_{n\times1},\,
\bm{r}(t)=\begin{pmatrix}r_{1}(t)\\
r_{2}(t)\\
\vdots\\
r_{p}(t)
\end{pmatrix}_{p\times1}\,
\end{equation}

\begin{equation}
\bm{r}=(\bm{r}_{1},\bm{r}_{2},...,\bm{r}_{p})_{n\times p}=\begin{pmatrix}\bm{r}(t)'\\
\bm{r}(t+1)'\\
\vdots\\
\bm{r}(T)'
\end{pmatrix}_{n\times p}, \,
\bm{R} = (\bm{R}_1, \bm{R}_2, ..., \bm{R}_N)
\label{eq: amf_vec_end}
\end{equation}
}Recall that we assume that the coefficients $\beta_{ij}$ are constants.
This assumption is only reasonable when the time period $(t,T)$ is
small, say three years, so the number of observations $n\approx150$
given we employ weekly data. Therefore, our sample size $n$ in this
regression is substantially less than the number of basis assets $p$.

We fit the GIBS algorithm to select the
basis assets set $S$ ($S$ is derived near the end). Then, the model
becomes
\begin{equation}
\bm{R}_{i}-\bm{r}_{0}=\alpha_{i}\vone_{n}+(\bm{r}_{S}
-(\bm{r}_{0})_{S}\vone'_{p})(\bm{\beta}_{i})_{S}+\bm{\epsilon}_{i}\quad.
\label{eq: amf_theory_ols_subset}
\end{equation}
Here, the intercept and the significance of each basis asset can be
tested, making the identifications $\bm{y}=\bm{R}_{i}-\bm{r}_{0}$
and $\bm{X}=\bm{r}-\bm{r_{0}}\vone_{p}'$ in Equation (\ref{eq: lasso}). Goodness of fit tests, comparisons of the in-sample adjusted $R^{2}$, and prediction out-of-sample $R^2$ \cite{campbell2008predicting} can be employed.

An example of Equation (\ref{eq: amf_theory_ols_subset}) is the Fama-French
(2015) \cite{fama2015five} five-factor model where all
of the basis assets are risk-factors, earning non-zero expected excess
returns. Here, the five traded risk-factors are: (i) the market portfolio
less the spot rate of interest ($R_{m}-R_{f}$), (ii) a portfolio
representing the performance of small (market capital) versus big
(market capital) companies ($SMB$), (iii) a portfolio representing
the performance of high book-to-market ratio versus small book-to-market
ratio companies ($HML$), (iv) a portfolio representing the performance
of robust (high) profit companies versus that of weak (low) profits
($RMW$), and (v) a portfolio representing the performance of firms
investing conservatively and those investing aggressively ($CMA$),
i.e.
\begin{equation}
\begin{array}{ll}
R_{i}(t)-r_{0}(t) =
& \alpha_{i}+\beta_{mi}(R_{m}(t)-r_{0}(t))+\beta_{si} SMB(t)+\beta_{hi} HML(t)\\
& + \beta_{ri} RMW(t)+\beta_{ci} CMA(t)+\epsilon_{i}(t).
\end{array}\label{eq: ff5}
\end{equation}
The key difference between the Fama-French five-factor and Equation
(\ref{eq: amf_theory_ols_subset}) is that Equation (\ref{eq: amf_theory_ols_subset}) allows distinct
securities to be related to different basis assets, many of which
may not be risk-factors, chosen from a larger set of basis assets
than just these five. In fact, we allow the number of basis assets
$p$ to be quite large (e.g. over one thousand), which enables the
number of non-zero coefficients $\bm{\beta}_{i}$ to be different
for different securities. As noted above, we also assume the coefficient
vector $\bm{\beta}_{i}$ to be sparse. The traditional literature,
which includes the Fama-French five-factor model, limits the regression
to a small number of risk-factors. In contrast, using the LASSO method,
we are able to fit our model using time series data when $p>n$, as
long as the $\beta_{i}$ coefficients are sparse and the basis assets
are not highly correlated. As noted previously, we handle this second
issue via clustering methods.

\section{The Estimation Procedure (GIBS algorithm)}

\label{sec: gibs_algo}

This section discusses the estimation procedure for the basis asset implied Adaptive Multi-Factor (AMF) model. To overcome the high-dimension and high-correlation difficulties, we propose a \textbf{Groupwise Interpretable Basis Selection (GIBS) algorithm} to empirically estimate the AMF model. The details are given in this section, and the sketch of the GIBS algorithm is shown in Table \ref{tab: gibs_algo} at the end of this section.

The data consists of security returns and all the ETFs available in
the CRSP database over the three year time period from January 2014
to December 2016. The same approach can be used in other time periods
as well. However, in earlier time periods, there were less ETFs. In
addition, in the collection of basis assets we include the five Fama-French
factors. A security is included in our sample only if it has prices
available for more than 80\% of all the trading weeks. For easy comparison,
companies are classified according to the first 2 digits of their
SIC code (a detailed description of SIC code classes can be found
in Appendix \ref{sec: amf_comp_classes}).

Suppose that we are given $p_{1}$ tradable basis assets $\bm{r}_{1},\bm{r}_{2},...,\bm{r}_{p_{1}}$.
In our investigation, these are returns on traded ETFs, and for comparison
to the literature, the Fama-French 5 factors. Using recent year data,
the number of ETFs is large, slightly over 1000 ($p_{1}\approx1000$).
Since these basis assets are highly-correlated, it is problematic
to fit $\bm{R}_{i}-\bm{r}_{0}$ directly on these basis assets using
a LASSO regression. Hence, we use the Prototype Clustering method
discussed in Section \ref{sec: proto_lasso} to reduce the number
of basis assets by selecting low-correlated representatives. Then,
we fit a modified version of the LASSO regression to these low-correlated representatives.
This improves the fitting accuracy and also selects a sparser and
more interpretable model.

For notation simplicity, denote
\begin{equation}
  \bm{Y}_{i}=\bm{R}_{i}-\bm{r}_{0}\,, \quad \bm{X}_{i}=\bm{r}_{i}-\bm{r}_{0}\,, \quad
  \bm{Y}=\bm{R}-\bm{r}_{0}\,, \quad \bm{X}=\bm{r}-\bm{r}_{0}
\end{equation}
where the definition of $\bm{R}_i, \, \bm{R}, \, \bm{r}_i, \, \bm{r}$ are in equation (\ref{eq: amf_vec} - \ref{eq: amf_vec_end}). Let $\bm{r}_{1}$ denote the market return. It is easy to check that most of the ETF basis assets $\bm{X}_{i}$ are correlated with $\bm{X}_{1}$ (the market return minus the risk free rate). We note that this pattern is not true for the other four Fama-French factors. Therefore, we
first orthogonalize every other basis asset to $\bm{X}_{1}$ before
doing the clustering and the LASSO regression. By orthogonalizing with
respect to the market return, we apvoid choosing redundant basis assets
similar to it and meanwhile, increase the accuracy of fitting. Note
that for OLS, projection does not affect the estimation since it only
affects the coefficients, not the estimated $\bm{\hat{y}}$. However,
in the LASSO, projection does affect the set of selected basis assets
because it changes the magnitude of shrinking.
Thus, we compute
\begin{equation}
\bm{\widetilde{X}}_{i}=(\bm{I}-P_{\bm{X}_{1}})\bm{X}_{i}
=(\bm{I}-\bm{X}_{1}(\bm{X}'_{1}\bm{X}_{1})^{-1}\bm{X}_{1}')\bm{X}_{i}
\qquad\text{where}\quad2\leq i\leq p_{1}
\label{eq: proj1}
\end{equation}
where $P_{\bm{X}_{1}}$ denotes the projection operator. Denote the
vector
\begin{equation}
\bm{\widetilde{X}}=(\bm{X}_{1},\bm{\widetilde{X}}_{2},\bm{\widetilde{X}}_{3},...,
\bm{\widetilde{X}}_{p_{1}}).
\label{eq: proj2}
\end{equation}
Note that this is equivalent to the residuals after regressing other
basis assets on the market return minus the risk free rate.

The transformed ETF basis assets $\bm{\widetilde{X}}$ contain highly
correlated members. We first divide these basis assets into categories
$A_{1},A_{2},...,A_{k}$ based on their financial interpretation. Note that $A\equiv\cup_{i=1}^{k}A_{i}=\{1,2,...,p_{1}\}$. The list of categories with more descriptions can be found in Appendix \ref{ap: amf_etf_classes}. The categories are (1) bond/fixed income, (2) commodity, (3) currency, (4) diversified portfolio, (5) equity, (6) alternative ETFs, (7) inverse, (8) leveraged, (9) real estate, and (10) volatility.

Next, from each category we need to choose a set of representatives.
These representatives should span the categories they are from, but
also have low correlation with each other. This can be done by using
the prototype-clustering method with distance defined by equation
(\ref{eq: corr_dist}), which yield the ``prototypes'' (representatives) within each cluster (intuitively, the prototype is at the center of each cluster) with low-correlations.

Within each category, we use the prototype clustering methods previously
discussed to find the set of representatives. The number of representatives
in each category can be decided according to a correlation threshold.
(Alternatively, we can also use the PCA dimension or other parameter
tuning methods to decide the number of prototypes. Note that even
if we use the PCA dimension to suggest the number of prototypes to
keep, the GIBS algorithm does not use any linear combinations of factors
as PCA does). This gives the sets $B_{1},B_{2},...,B_{k}$ with $B_{i}\subset A_{i}$
for $1\leq i\leq k$. Denote $B\equiv\cup_{i=1}^{k}B_{i}$. Although
this reduction procedure guarantees low-correlation between the elements
in each $B_{i}$, it does not guarantee low-correlation across the
elements in the union $B$. So, an additional step is needed, which
is prototype clustering on $B$ to find a low-correlated representatives
set $U$. Note that $U\subseteq B$. Denote $p_{2}\equiv\#U$. The list
of all ETFs in the set $U$ is given in Appendix \ref{ap: etf_representatives} Table \ref{tab: amf_etf_representatives}. This is still a large set with $p_{2}=186$.

Recall from the notation Section \ref{sec: notations} that $\bm{\w{X}}_{U}$ means the columns of the matrix $\bm{\w{X}}$ indexed by the set $U$. Since basis assets in $\bm{\widetilde{X}}_{U}$ are not highly correlated, a LASSO regression can be applied.  By equation (\ref{eq: lasso_beta}), we have that
\begin{equation}
\bm{\widetilde{\beta}}_{i}=\underset{\bm{\beta}_{i}\in\mathbb{R}^{p},(\bm{\beta}_{i})_{j}
=0(\forall j\in U^{c})}{\arg\min}\left\{ \frac{1}{2n}\norm{\bm{Y}_{i}-\bm{\widetilde{X}\beta}_{i}}_{2}^{2}
+\lambda\norm{\bm{\beta}_{i}}_{1}\right\}
\end{equation}
where $U^{c}$ denotes the complement of $U$. However, here we use
a different $\lambda$ compared to the traditional LASSO. Normally
the $\lambda$ of LASSO is selected by the cross-validation. However
this will overfit the data as discussed in Section \ref{sec: amf_comparison}.
So here we use a modified version of the $\lambda$ select rule and
set
\begin{equation}
  \lambda=\max\{\lambda_{1se},min\{\lambda:\#supp(\bm{\widetilde{\beta}}_{i})\leq20\}\}
\end{equation}
where $\lambda_{1se}$ is the $\lambda$ selected by the ``1se rule''. The ``1se rule'' gives the most regularized model such that error is within one standard error of the minimum error achieved by the cross-validation (see \cite{friedman2010regularization, simon2011Regularization, tibshirani2012strong}). Further discussion of the choice of $\lambda$ can be found in Section \ref{sec: amf_comparison}.

The set of basis assets selected is
\begin{equation}
S_i \equiv supp(\bm{\widetilde{\beta}}_{i})
\label{eq: factor_select}
\end{equation}
Next, we fit an Ordinary Least Squares (OLS) regression on the selected basis assets. Since this
is an OLS regression, we use the original basis assets $\bm{X}_{S_{i}}$
rather than the orthogonalized basis assets with respect to the market
return $\bm{\widetilde{X}}_{S_i}$. In this way, we construct the set of basis assets $S_{i}$.

Note that here we can also add the Fama-French 5 factors into $S_{i}$ if not selected, which will be also discussed in Section \ref{sec: amf_comparison} as the GIBS + FF5 model. This is included to compare our results with the literature. However, the comparison results between the GIBS and the GIBS + FF5 model in Section \ref{sec: amf_comparison} show that adding back Fama-French 5 factors into $S_{i}$ results in overfitting and should be avoided. Hence, the GIBS algorithm emplyed herein doesn't include the Fama-French 5 factors if they are not selected in the procedure above.
%%_{i}\ensuremath{}\ensuremath{}\ensuremath{}\ensuremath{}\ensuremath{}

%Note that after the above procedure there might be some risk-factors
%in the set $\widetilde{S}'$ still correlated with the market return
%($\bm{X}_{1}$) because of that we projected out the market return
%before removing the high-correlated columns and that we pick about
%20 risk-factors for each company with LASSO (some weak risk-factors
%may still be selected if there are not enough strong risk-factors.)
%Therefore, we need to remove these weak risk-factors, in other word,
%removing risk-factors which are still highly correlated to $\bm{X}_{1}$
%from the set $\widetilde{S}_{i}'$.

The following OLS regression is used to estimate $\bm{\hat\beta}_{i}$,
the OLS estimator of $\bm{\beta}_{i}$ in
\begin{equation}
\bm{Y}_{i}=\alpha_{i}\vone_{n}
+\bm{X}_{S_{i}}(\bm{\beta}_{i})_{S_{i}}
+\bm{\epsilon}_{i}.
\label{eq: amf_ols_subset}
\end{equation}
Note that $supp(\bm{\hat\beta}_{i})\subseteq {S}_{i}$. The
adjusted $R^{2}$ is obtained from this estimation. Since we are in the
OLS regime, significance tests can be performed on $\bm{\hat\beta}_{i}$.
This yields the significant set of coefficients
\begin{equation}
S_{i}^{*}\equiv\{j:P_{H_{0}}(|\bm{\beta}_{i,j}|\geq|\bm{\hat\beta}_{i,j}|)<0.05\}
\quad\text{where}\quad H_{0}:\text{True value}\,\,\bm{\beta}_{i,j}=0.
\label{eq: factor_sgn}
\end{equation}

Note that the significant basis asset set is a subset of the selected basis asset set. In other words,
\begin{equation}
S_i^* \subseteq supp(\bm{\hat\beta}_i) \subseteq S_i \subseteq \{1, 2, ..., p\} .
\end{equation}

To sum up, the sketch of the GIBS algorithm is shown in Table \ref{tab: gibs_algo}. Recall from the notation Section \ref{sec: notations} that for an index set $ S \subseteq \{1, 2, ..., p\} $, $\bm{\w{X}}_{S}$ means the columns of the matrix $\bm{\w{X}}$ indexed by the set $S$.

Some follow-up tests and applications of the GIBS algorithm we propose here can be found in \cite{zhu2021time, zhu2021news, jarrow2021low}. Some related works can be found in \cite{zhu2021clustering, zhu2020adaptive, li2021frequentnet, huang2021staying, huang2020time, jie2018stochastic, zhao2018examining, prashanth2016cumulative, du2020multiple, zhang2021form, mao2020differentiate, mao2020speculative, stein2021quclassi, stein2020hybrid, stein2021qugan, du2021improved, li2021two, davis2021clustering, wang2020efficient, zhao2020missing, zhao2020matrix, bo2021subspace}.

\begin{table}[H]
\center
\begin{tabular}{|| l ||}
\hhline{|=|}
\\[-1.6ex]
\textbf{The Groupwise Interpretable Basis Selection (GIBS) algorithm }  \\[1ex]
\hhline{|=|}
\\[-1.6ex]
Inputs: Stocks to fit $\bm{Y}$ and basis assets $\bm{X}$.
\\[0.6ex]
\hline
\\[-1.6ex]
1. Derive $\bm{\widetilde{X}}$ using $\bm{X}$ and the equation (\ref{eq: proj1}, \ref{eq: proj2}).
\\[0.6ex]
2. Divide the transformed basis assets $\bm{\widetilde{X}}$ into k groups $A_{1},A_{2},\cdots A_{k}$ by a
\\[0.6ex]
\hspace{0.4cm} financial interpretation.
\\[0.6ex]
3. Within each group, use prototype clustering to find prototypes $ B_i \subset A_i $.
\\[0.6ex]
4. Let $ B = \cup_{i = 1}^k B_i $, use prototype clustering in B to find prototypes $ U \subset B $.
\\[0.6ex]
5. For each stock $\bm{Y}_{i}$, use a modified version of LASSO to reduce $\bm{\widetilde{X}}_{U}$
\\[0.6ex]
\hspace{0.4cm} to the selected basis assets $\bm{\widetilde{X}}_{S_{i}}$
\\[0.6ex]
6. For each stock $ \bm{Y}_i $, fit linear regression on $ \bm{X}_{S_i} $.
\\[0.6ex]
\hline
\\[-1.6ex]
Outputs: Selected factors $ S_i $, significant factors $ S_i^*$, and coefficients in step 6.
\\[0.6ex]
\hhline{|=|}
\end{tabular}
\caption{The sketch of Groupwise Interpretable Basis Selection (GIBS) algorithm}
\label{tab: gibs_algo}
\end{table}

It is also important to understand which basis assets affect which
securities. Given the set of securities is quite large, it is more
reasonable to study which classes of basis assets affect which classes
of securities. The classes of basis assets are given in Appendix \ref{ap: amf_etf_classes},
and the classes of securities classified by the first 2 digits of
their SIC code are in Appendix \ref{sec: amf_comp_classes}.
For each security class, we count the number of significant basis
asset classes as follows.

Recall that $N$ is the number of securities. Denote $l$ to be the
number of security classes. Denote the security classes by $C_{1},C_{2},C_{3},...,C_{l}$
where $\bigcup_{d=1}^{l}C_{d}=\{1,2,...,N\}$. Recall that the number
of basis assets is $p$. Let the number of basis asset classes be
$m$. Let the basis asset classes be denoted $F_{1},F_{2},F_{3},...,F_{m}$
where $\bigcup_{b=1}^{m}F_{b}=\{1,2,...,p\}$ and $p$ is the number
of basis assets which were significant for at least one of the security
i. Also recall that $ S_i^* \subseteq \{1, 2, ..., p\} $ in equation (\ref{eq: factor_sgn}). Denote the significant count matrix to be $\bm{A}=\{a_{b,d}\}_{m\times l}$
where
\begin{equation}
a_{b,d}=\sum_{i\in C_{d}} \#\{ S_i^* \cap F_{b} \}.
\end{equation}
That is, each element $a_{b,d}$ of matrix $\bm{A}$ is the number
of significant basis assets in basis asset class $b$, selected by
securities in class $d$. Finally, denote the proportion matrix to
be $\bm{G}=\{g_{b,d}\}_{m\times l}$ where
\begin{equation}
g_{b,d}=\frac{a_{b,d}}{\sum_{1\leq j\leq m}a_{j,d}}.\label{eq: perc_table}
\end{equation}
In other words, each element $g_{b,d}$ of matrix $\bm{G}$ is the
proportion of significant basis assets in basis asset class $b$ selected
by security class $d$ among all basis assets selected by security
class $d$. Note that the elements in each column of $\bm{G}$ sum
to one.

\section{Estimation Results}

\label{sec: amf_results}

Our results show that the GIBS algorithm selects a total of 186
basis assets from different sectors for at least one company. And
all of these 186 basis assets are significant in the second stage
OLS regressions after the GIBS selection for at least one company.
This validates our assumption that the total number of basis assets
is large; much larger than 10 basis assets, which is typically the
maximum number of basis asset with non-zero risk premiums (risk-factors)
seen in the literature (see Harvey, Liu, Zhu (2016) \cite{harvey2016and}).

\begin{figure}[H]
\center \vspace{0in}
\includegraphics[width=0.8\textwidth]{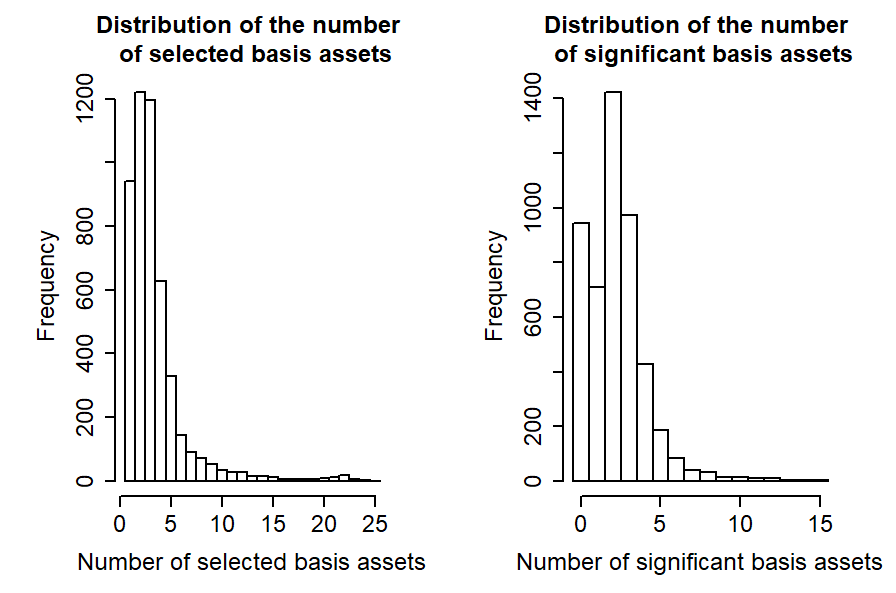}
\caption{Distribution of the number of basis assets. The left figure shows the histogram of the number of basis assets selected by GIBS. On the right we report the histogram of number of basis assets significant in the second-step OLS regression at 5\% level.}
\label{fig: amf_n_factors} \vspace{0in}
\end{figure}

In addition, the results validate our sparsity assumption, that each
company is significantly related to only a small number of basis assets
(at a 5\% level of significance). Indeed, for each company an average
of 3.4 basis assets are selected by GIBS and an average of 2.2 basis assets show significance in the second stage OLS regression. (Even using the traditional cross-validation method with overfitting discussed in Section \ref{sec: amf_comparison}, only an average of 13.4 basis assets are selected.) In other words, the average number of elements in $S_i$ (see Equation \ref{eq: factor_select}) is 3.4 and the average number of elements in $S_{i}$ (see Equation \ref{eq: factor_sgn}) is 2.2. Figure \ref{fig: amf_n_factors}
shows the distribution of the number of basis assets selected by GIBS and the number of basis assets that are significant in the second stage OLS regression. As depicted, most securities have between $1\sim10$ significant basis assets. Thus high dimensional methods are appropriate and necessary here.

\begin{figure}[H]
\centering
\includegraphics[width=\textwidth]{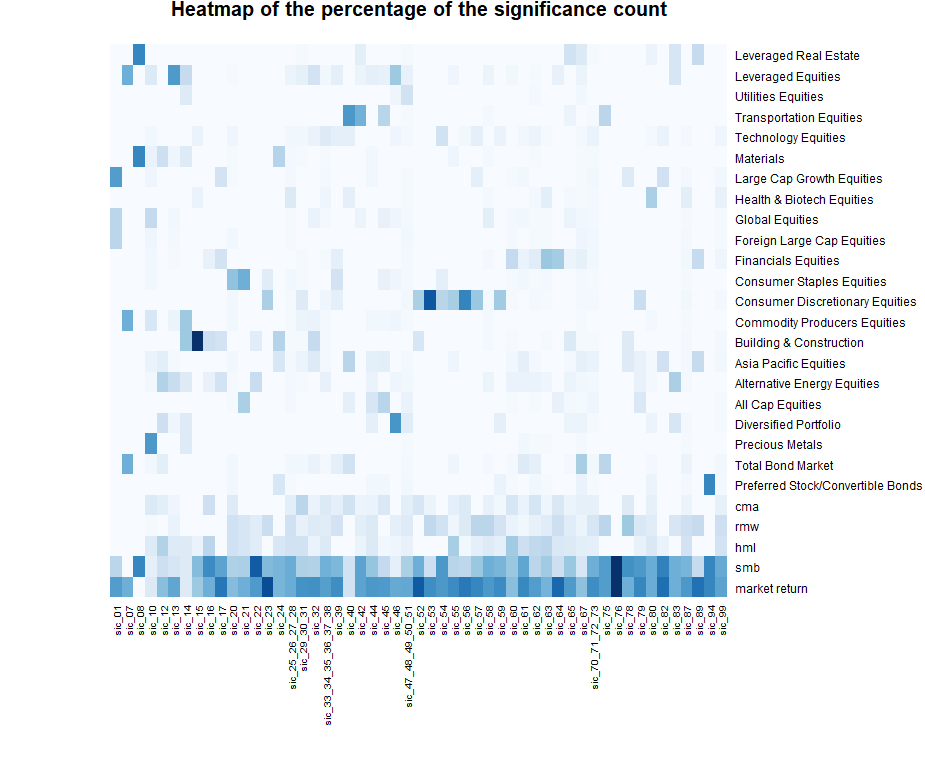}
\caption{Heat map of percentage of significance count. Each grid are related
to the percentage of basis assets selected in the corresponding sector
(shown as the row name) by the company group (classified by the SIC
code shown as the column name). The figure shows that different company
groups may choose some basis assets in common, but also tends to select
different sectors of basis assets.}
\label{fig: amf_heatmap}
\end{figure}

Table \ref{tab: amf_perc_sgn} provides the matrix $\bm{G}$ in percentage. Each
grid is $100\cdot g_{b,d}$ where $g_{b,d}$ is defined in equation
(\ref{eq: perc_table}). Figure \ref{fig: amf_heatmap} is a heat map from which we can
visualize patterns in Table \ref{tab: amf_perc_sgn}. The darker the grid, the
larger the percentage of significant basis assets. As indicated, different
security classes depend on different classes of basis assets, although
some basis assets seem to be shared in common. Not all of the Fama-French
5 risk-factors are significant in presence of the additional basis
assets in our model. Only the market portfolio shows a strong significance
for nearly all securities. The emerging market equities and the money
market ETF basis assets seem to affect many securities as well. As
shown, all of the basis assets are needed to explain security returns
and different securities are related to a small number of different
basis assets.

\begin{table}[H]
%\resizebox{6in}{!}{ %\vspace*{-1cm} \hspace*{-1cm}
%\resizebox{!}{0.49\textwidth}{
%\begin{minipage}{0.95\textwidth}
\centering
\begin{adjustbox}{width=\textwidth, totalheight=\textheight,keepaspectratio}
\begin{tabular}{|| *{19}{c|} |}
\hhline{|*{19}{=}|}
\multirow{2}{*}{ETF Class}  & \multicolumn{18}{c||}{SIC First 2 digits}\tabularnewline
\multirow{2}{*}{} & 01 & 07 & 08 & 10 & 12 & 13 & 14 & 15 & 16 & 17 & 20 & 21 & 22 & 23 & 24 & 25-28 & 29-31 & 32 \\
\hhline{|*{19}{=|}|}
market return & 29 & 25 & 0 & 7 & 21 & 27 & 6 & 19 & 25 & 36 & 22 & 25 & 29 & 44 & 23 & 25 & 31 & 31 \\
\hline
smb & 14 & 0 & 33 & 5 & 11 & 8 & 6 & 22 & 32 & 27 & 16 & 17 & 41 & 22 & 23 & 25 & 16 & 16 \\
\hline
hml & 0 & 0 & 0 & 6 & 16 & 6 & 6 & 3 & 14 & 0 & 10 & 8 & 6 & 0 & 8 & 10 & 10 & 3 \\
\hline
rmw & 0 & 0 & 0 & 1 & 0 & 3 & 0 & 0 & 0 & 0 & 10 & 8 & 6 & 11 & 0 & 10 & 4 & 6 \\
\hline
cma & 0 & 0 & 0 & 7 & 5 & 2 & 0 & 0 & 11 & 0 & 6 & 0 & 0 & 0 & 0 & 5 & 14 & 3 \\
\hline
Pref. Stock/Convrtbl. Bonds & 0 & 0 & 0 & 0 & 0 & 0 & 0 & 0 & 0 & 0 & 2 & 0 & 0 & 0 & 8 & 1 & 0 & 0 \\
\hline
Total Bond Market & 0 & 25 & 0 & 2 & 5 & 0 & 0 & 0 & 0 & 0 & 2 & 0 & 0 & 0 & 0 & 2 & 2 & 3 \\
\hline
Precious Metals & 0 & 0 & 0 & 30 & 0 & 0 & 6 & 0 & 0 & 0 & 0 & 0 & 0 & 0 & 0 & 0 & 0 & 0 \\
\hline
Diversified Portfolio & 0 & 0 & 0 & 0 & 11 & 2 & 6 & 0 & 0 & 0 & 0 & 0 & 0 & 0 & 0 & 0 & 0 & 0 \\
\hline
All Cap Equities & 0 & 0 & 0 & 0 & 0 & 0 & 0 & 0 & 0 & 0 & 1 & 17 & 0 & 0 & 0 & 1 & 0 & 0 \\
\hline
Alternative Energy Equities & 0 & 0 & 0 & 2 & 16 & 12 & 6 & 0 & 4 & 9 & 1 & 0 & 12 & 0 & 0 & 1 & 1 & 0 \\
\hline
Asia Pacific Equities & 0 & 0 & 0 & 3 & 5 & 1 & 0 & 0 & 0 & 0 & 2 & 0 & 0 & 0 & 8 & 1 & 3 & 6 \\
\hline
Building \& Construction & 0 & 0 & 0 & 0 & 0 & 0 & 19 & 50 & 11 & 9 & 0 & 0 & 6 & 0 & 15 & 1 & 1 & 12 \\
\hline
Commodity Producers Equities & 0 & 25 & 0 & 8 & 0 & 3 & 19 & 0 & 0 & 0 & 2 & 0 & 0 & 0 & 0 & 1 & 3 & 3 \\
\hline
Consumer Discrtnry. Equities & 0 & 0 & 0 & 0 & 0 & 0 & 0 & 0 & 0 & 0 & 2 & 0 & 0 & 17 & 0 & 1 & 6 & 0 \\
\hline
Consumer Staples Equities & 0 & 0 & 0 & 2 & 0 & 0 & 0 & 0 & 0 & 0 & 20 & 25 & 0 & 6 & 0 & 2 & 1 & 0 \\
\hline
Financials Equities & 0 & 0 & 0 & 2 & 0 & 1 & 0 & 0 & 4 & 9 & 0 & 0 & 0 & 0 & 0 & 0 & 1 & 0 \\
\hline
Foreign Large Cap Equities & 14 & 0 & 0 & 2 & 0 & 1 & 0 & 0 & 0 & 0 & 2 & 0 & 0 & 0 & 0 & 1 & 0 & 0 \\
\hline
Global Equities & 14 & 0 & 0 & 13 & 0 & 2 & 0 & 0 & 0 & 0 & 0 & 0 & 0 & 0 & 0 & 1 & 0 & 3 \\
\hline
Health \& Biotech Equities & 0 & 0 & 0 & 0 & 0 & 0 & 0 & 3 & 0 & 0 & 0 & 0 & 0 & 0 & 0 & 7 & 0 & 0 \\
\hline
Large Cap Growth Equities & 29 & 0 & 0 & 2 & 0 & 0 & 0 & 0 & 0 & 9 & 1 & 0 & 0 & 0 & 0 & 1 & 0 & 0 \\
\hline
Materials & 0 & 0 & 33 & 3 & 11 & 2 & 6 & 0 & 0 & 0 & 1 & 0 & 0 & 0 & 15 & 1 & 1 & 0 \\
\hline
Technology Equities & 0 & 0 & 0 & 2 & 0 & 0 & 0 & 3 & 0 & 0 & 2 & 0 & 0 & 0 & 0 & 2 & 2 & 3 \\
\hline
Transportation Equities & 0 & 0 & 0 & 0 & 0 & 0 & 0 & 0 & 0 & 0 & 0 & 0 & 0 & 0 & 0 & 0 & 0 & 0 \\
\hline
Utilities Equities & 0 & 0 & 0 & 0 & 0 & 0 & 6 & 0 & 0 & 0 & 0 & 0 & 0 & 0 & 0 & 0 & 0 & 0 \\
\hline
Leveraged Equities & 0 & 25 & 0 & 7 & 0 & 29 & 12 & 0 & 0 & 0 & 1 & 0 & 0 & 0 & 0 & 3 & 4 & 9 \\
\hline
Leveraged Real Estate & 0 & 0 & 33 & 0 & 0 & 0 & 0 & 0 & 0 & 0 & 0 & 0 & 0 & 0 & 0 & 0 & 1 & 0 \\
\hhline{|*{19}{=}|}
\end{tabular}
%\end{minipage}
%}

\end{adjustbox}
\caption{Table for the percentage of significance count. The table provides the matrix $\bm{G}$ in percentage (each grid is $100\cdot g_{b,d}$ where $g_{b,d}$ is defined in equation (\ref{eq: perc_table})). Each grid is the percentage of the basis asset selected in the corresponding sector (shown as the row name) by the company group (classified by the SIC
code shown as the column name). Note that the elements in each column add up to 100, which means 100\% (maybe be slightly different from 100 due to the rounding issue). The percent signs are omitted to save space.}
\label{tab: amf_perc_sgn}
\end{table}

%\hspace*{-1cm}
\begin{table}[H]
%\resizebox{!}{0.5\textwidth}{
%\begin{minipage}{0.95\textwidth}
\centering
\begin{adjustbox}{width=\textwidth, totalheight=\textheight,keepaspectratio}
\begin{tabular}{|| *{19}{c|} |}
\hhline{|*{19}{=}|}
\multirow{2}{*}{ETF Class}  & \multicolumn{18}{c||}{SIC First 2 digits}\tabularnewline
\multirow{2}{*}{} & 33-38 & 39 & 40 & 42 & 44 & 45 & 46 & 47-51 & 52 & 53 & 54 & 55 & 56 & 57 & 58 & 59 & 60 & 61 \\
\hhline{|*{19}{=|}|}
\hline
market return & 28 & 32 & 10 & 27 & 30 & 29 & 26 & 26 & 42 & 32 & 30 & 32 & 36 & 33 & 30 & 32 & 21 & 33 \\
\hline
smb & 24 & 23 & 10 & 27 & 20 & 13 & 5 & 17 & 25 & 11 & 30 & 15 & 14 & 19 & 25 & 23 & 19 & 27 \\
\hline
hml & 7 & 0 & 0 & 3 & 7 & 2 & 2 & 4 & 0 & 0 & 0 & 18 & 2 & 5 & 6 & 5 & 18 & 11 \\
\hline
rmw & 6 & 9 & 0 & 5 & 7 & 0 & 5 & 6 & 0 & 13 & 10 & 2 & 7 & 14 & 14 & 9 & 3 & 2 \\
\hline
cma & 7 & 5 & 10 & 3 & 3 & 7 & 0 & 5 & 8 & 3 & 0 & 5 & 7 & 0 & 4 & 2 & 8 & 3 \\
\hline
Pref. Stock/Convrtbl. Bonds & 0 & 0 & 0 & 0 & 2 & 2 & 0 & 2 & 0 & 0 & 0 & 0 & 0 & 0 & 0 & 3 & 1 & 0 \\
\hline
Total Bond Market & 1 & 5 & 10 & 0 & 0 & 0 & 0 & 1 & 0 & 0 & 0 & 2 & 0 & 0 & 1 & 0 & 1 & 6 \\
\hline
Precious Metals & 0 & 0 & 0 & 0 & 0 & 0 & 0 & 1 & 0 & 0 & 0 & 0 & 0 & 0 & 0 & 0 & 0 & 2 \\
\hline
Diversified Portfolio & 0 & 0 & 0 & 0 & 5 & 0 & 30 & 6 & 0 & 0 & 0 & 0 & 0 & 0 & 0 & 5 & 0 & 0 \\
\hline
All Cap Equities & 0 & 0 & 5 & 0 & 8 & 15 & 0 & 3 & 0 & 0 & 0 & 0 & 0 & 0 & 0 & 0 & 1 & 0 \\
\hline
Alternative Energy Equities & 4 & 0 & 0 & 0 & 3 & 0 & 2 & 2 & 0 & 0 & 0 & 0 & 0 & 0 & 3 & 0 & 4 & 3 \\
\hline
Asia Pacific Equities & 1 & 0 & 15 & 0 & 5 & 5 & 0 & 1 & 0 & 0 & 0 & 2 & 0 & 0 & 0 & 0 & 2 & 5 \\
\hline
Building \& Construction & 1 & 0 & 0 & 0 & 0 & 0 & 0 & 0 & 8 & 0 & 0 & 0 & 0 & 0 & 1 & 0 & 1 & 0 \\
\hline
Commodity Producers Equities & 1 & 0 & 0 & 0 & 2 & 2 & 2 & 1 & 0 & 0 & 0 & 0 & 0 & 0 & 1 & 0 & 0 & 0 \\
\hline
Consumer Discrtnry. Equities & 2 & 5 & 0 & 0 & 0 & 0 & 0 & 2 & 17 & 42 & 15 & 18 & 34 & 19 & 1 & 18 & 1 & 0 \\
\hline
Consumer Staples Equities & 0 & 9 & 0 & 0 & 0 & 4 & 2 & 1 & 0 & 0 & 5 & 0 & 0 & 0 & 0 & 0 & 1 & 2 \\
\hline
Financials Equities & 1 & 5 & 0 & 0 & 0 & 0 & 0 & 1 & 0 & 0 & 0 & 0 & 0 & 0 & 1 & 0 & 13 & 3 \\
\hline
Foreign Large Cap Equities & 0 & 0 & 0 & 0 & 0 & 0 & 0 & 2 & 0 & 0 & 0 & 0 & 0 & 0 & 0 & 0 & 2 & 0 \\
\hline
Global Equities & 1 & 0 & 0 & 3 & 0 & 4 & 2 & 1 & 0 & 0 & 0 & 0 & 0 & 0 & 5 & 0 & 1 & 2 \\
\hline
Health \& Biotech Equities & 2 & 0 & 5 & 0 & 0 & 0 & 0 & 1 & 0 & 0 & 0 & 0 & 0 & 0 & 1 & 0 & 0 & 0 \\
\hline
Large Cap Growth Equities & 1 & 0 & 0 & 0 & 2 & 0 & 0 & 1 & 0 & 0 & 0 & 0 & 0 & 5 & 0 & 0 & 1 & 2 \\
\hline
Materials & 1 & 0 & 0 & 0 & 2 & 0 & 0 & 0 & 0 & 0 & 0 & 2 & 0 & 0 & 1 & 0 & 0 & 0 \\
\hline
Technology Equities & 6 & 5 & 5 & 0 & 0 & 0 & 2 & 1 & 0 & 0 & 10 & 0 & 2 & 5 & 0 & 3 & 0 & 2 \\
\hline
Transportation Equities & 0 & 0 & 30 & 24 & 0 & 15 & 0 & 1 & 0 & 0 & 0 & 0 & 0 & 0 & 1 & 0 & 0 & 0 \\
\hline
Utilities Equities & 0 & 0 & 0 & 0 & 0 & 0 & 2 & 10 & 0 & 0 & 0 & 0 & 0 & 0 & 0 & 0 & 0 & 0 \\
\hline
Leveraged Equities & 2 & 5 & 0 & 3 & 5 & 4 & 19 & 4 & 0 & 0 & 0 & 2 & 0 & 0 & 3 & 0 & 2 & 0 \\
\hline
Leveraged Real Estate & 0 & 0 & 0 & 5 & 0 & 0 & 0 & 1 & 0 & 0 & 0 & 0 & 0 & 0 & 0 & 0 & 0 & 0 \\
\hhline{|*{19}{=}|}
\end{tabular}
%\end{minipage}
%}
\end{adjustbox}
\caption*{Table \ref{tab: amf_perc_sgn} continued.}
\end{table}

%\hspace*{-1cm}
\begin{table}[H]
\centering
\begin{adjustbox}{width=\textwidth, totalheight=\textheight,keepaspectratio}
\begin{tabular}{|| *{18}{c|} |}
\hhline{|*{18}{=}|}
\multirow{2}{*}{ETF Class}  & \multicolumn{17}{c||}{SIC First 2 digits}\tabularnewline
\multirow{2}{*}{}  & 62 & 63 & 64 & 65 & 67 & 70-73 & 75 & 76 & 78 & 79 & 80 & 82 & 83 & 87 & 89 & 94 & 99 \\
\hhline{|*{18}{=}|}
& 62 & 63 & 64 & 65 & 67 & 70 & 75 & 76 & 78 & 79 & 80 & 82 & 83 & 87 & 89 & 94 & 99 \\
\hline
market return & 29 & 24 & 39 & 29 & 21 & 32 & 29 & 50 & 25 & 33 & 25 & 38 & 25 & 30 & 38 & 33 & 27 \\
\hline
smb & 14 & 18 & 11 & 19 & 8 & 25 & 29 & 50 & 31 & 30 & 25 & 34 & 25 & 23 & 12 & 33 & 26 \\
\hline
hml & 13 & 14 & 7 & 7 & 5 & 5 & 0 & 0 & 0 & 4 & 7 & 3 & 0 & 9 & 0 & 0 & 9 \\
\hline
rmw & 4 & 6 & 11 & 6 & 3 & 8 & 14 & 0 & 19 & 7 & 6 & 0 & 8 & 11 & 12 & 0 & 11 \\
\hline
cma & 9 & 2 & 7 & 3 & 4 & 4 & 0 & 0 & 6 & 0 & 4 & 0 & 0 & 3 & 0 & 0 & 4 \\
\hline
Pref. Stock/Convrtbl. Bonds & 2 & 1 & 0 & 0 & 3 & 1 & 0 & 0 & 0 & 0 & 3 & 0 & 0 & 1 & 0 & 33 & 1 \\
\hline
Total Bond Market & 5 & 0 & 0 & 1 & 17 & 2 & 14 & 0 & 0 & 0 & 1 & 0 & 0 & 3 & 0 & 0 & 2 \\
\hline
Precious Metals & 1 & 1 & 0 & 0 & 1 & 0 & 0 & 0 & 0 & 0 & 0 & 0 & 0 & 1 & 0 & 0 & 0 \\
\hline
Diversified Portfolio & 0 & 0 & 0 & 3 & 3 & 1 & 0 & 0 & 0 & 0 & 3 & 0 & 8 & 1 & 0 & 0 & 1 \\
\hline
All Cap Equities & 1 & 0 & 4 & 1 & 1 & 2 & 0 & 0 & 0 & 7 & 0 & 0 & 0 & 0 & 0 & 0 & 1 \\
\hline
Alternative Energy Equities & 3 & 2 & 0 & 3 & 2 & 1 & 0 & 0 & 0 & 4 & 1 & 0 & 17 & 1 & 0 & 0 & 0 \\
\hline
Asia Pacific Equities & 1 & 0 & 0 & 1 & 4 & 3 & 0 & 0 & 6 & 4 & 1 & 10 & 0 & 1 & 12 & 0 & 2 \\
\hline
Building \& Construction & 0 & 1 & 0 & 7 & 1 & 1 & 0 & 0 & 6 & 0 & 0 & 0 & 0 & 1 & 0 & 0 & 1 \\
\hline
Commodity Producers Equities & 0 & 0 & 0 & 0 & 1 & 0 & 0 & 0 & 0 & 0 & 1 & 0 & 0 & 1 & 0 & 0 & 0 \\
\hline
Consumer Discrtnry. Equities & 1 & 1 & 0 & 0 & 1 & 1 & 0 & 0 & 0 & 11 & 0 & 0 & 0 & 2 & 0 & 0 & 1 \\
\hline
Consumer Staples Equities & 1 & 1 & 0 & 0 & 1 & 1 & 0 & 0 & 0 & 0 & 0 & 0 & 0 & 1 & 0 & 0 & 0 \\
\hline
Financials Equities & 6 & 19 & 18 & 3 & 4 & 2 & 0 & 0 & 0 & 0 & 0 & 0 & 0 & 2 & 12 & 0 & 2 \\
\hline
Foreign Large Cap Equities & 1 & 1 & 0 & 0 & 2 & 2 & 0 & 0 & 0 & 0 & 0 & 0 & 0 & 1 & 0 & 0 & 1 \\
\hline
Global Equities & 1 & 1 & 0 & 3 & 2 & 1 & 0 & 0 & 0 & 0 & 0 & 0 & 0 & 1 & 0 & 0 & 1 \\
\hline
Health \& Biotech Equities & 1 & 5 & 0 & 1 & 1 & 1 & 0 & 0 & 0 & 0 & 17 & 0 & 0 & 5 & 0 & 0 & 4 \\
\hline
Large Cap Growth Equities & 0 & 0 & 0 & 0 & 1 & 2 & 0 & 0 & 6 & 0 & 0 & 10 & 0 & 1 & 0 & 0 & 2 \\
\hline
Materials & 1 & 0 & 0 & 0 & 1 & 1 & 0 & 0 & 0 & 0 & 0 & 0 & 0 & 0 & 0 & 0 & 0 \\
\hline
Technology Equities & 3 & 1 & 0 & 0 & 1 & 3 & 0 & 0 & 0 & 0 & 1 & 3 & 0 & 1 & 0 & 0 & 2 \\
\hline
Transportation Equities & 0 & 0 & 0 & 3 & 0 & 1 & 14 & 0 & 0 & 0 & 0 & 0 & 0 & 0 & 0 & 0 & 0 \\
\hline
Utilities Equities & 1 & 0 & 0 & 0 & 2 & 0 & 0 & 0 & 0 & 0 & 0 & 0 & 0 & 0 & 0 & 0 & 0 \\
\hline
Leveraged Equities & 1 & 0 & 4 & 0 & 3 & 0 & 0 & 0 & 0 & 0 & 0 & 0 & 8 & 0 & 0 & 0 & 0 \\
\hline
Leveraged Real Estate & 0 & 0 & 0 & 10 & 7 & 1 & 0 & 0 & 0 & 0 & 3 & 0 & 8 & 0 & 12 & 0 & 0 \\
\hhline{|*{18}{=}|}
\end{tabular}
\end{adjustbox}
\caption*{Table \ref{tab: amf_perc_sgn} continued.}
\end{table}

%\clearpage
\subsection{Intercept Test}

\begin{figure}[H]
\center \vspace{0cm}
\includegraphics[width=0.9\textwidth]{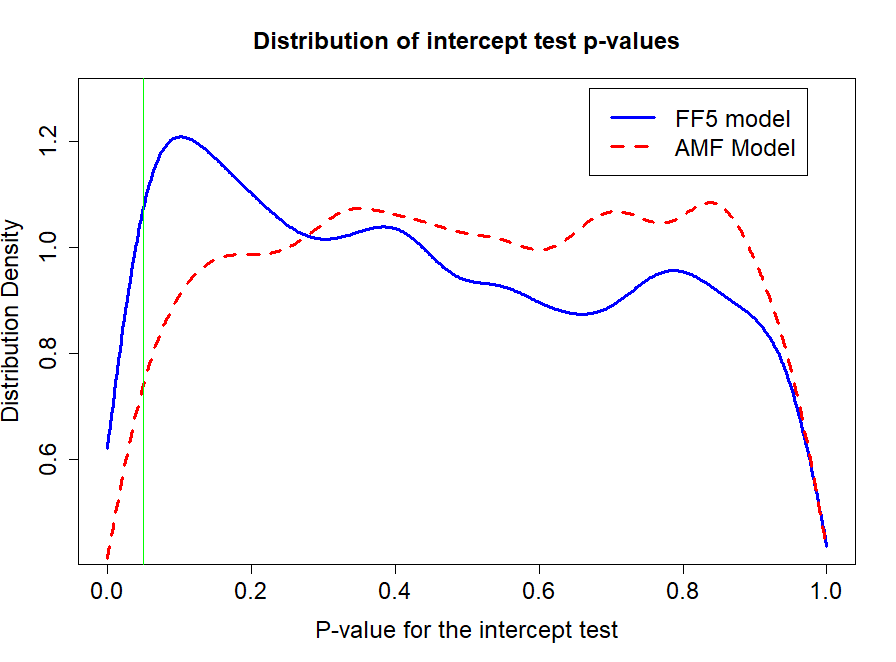}
\captionof{figure}{Comparison of intercept test p-values for the Fama-French 5-factor (FF5) model and the Adaptive Multi-factor (AMF) model.}
\label{fig: amf_int}
\end{figure}

This section provides the tests for a zero intercept. Using the Fama-French
5-factor model as a comparison, Figure \ref{fig: amf_int} compares the intercept
test p-values between our basis asset implied Adaptive Multi-factor (AMF) model and Fama-French 5-factor (FF5) model. As indicated, 6.33\% (above 5\%) of the securities have significant intercepts in the FF5 model, while 4.12\% (below 5\%) of the securities in AMF have significant $\alpha$'s. This may suggests that the AMF model is more insightful than the FF5 model, since AMF reveals more relevant factors and makes the intercept closer to 0.

Since we replicate this test for about 5000 stocks in the CRSP database,
it is important to control for a False Discovery Rate (FDR) because even
if there is a zero intercept, a replication of 5000 tests will have
about 5\% showing false significance. We adjust for the false discovery rate using the Benjamini-Hochberg (BH) procedure \cite{benjamini1995controlling}
and the Benjamini-Hochberg-Yekutieli (BHY) procedures \cite{benjamini2001control}.
The BH method does not account for the correlation between tests,
while the BHY method does. In our case, each test is done on an individual
stock, which may have correlations. So the BHY method is more appropriate
here.

Chordia et al. (2017) \cite{chordia2017p} suggests that the false
discovery proportion (FDP) approach in \cite{romano2007control} should
be applied rather than a false discovery rate procedure. The BH approach
only controls the expected value of FDP while FDP controls the family-wise
error rate directly. %We use the R package \textit{StepwiseTest} \cite{hsu2016stepwisetest}
%to perform the FDP test, which gives all zeros for the intercept test,
%meaning that no true discovery (significant non-zero intercept) was
%found. In other word, in our analysis, FDP gives the same results
%as BHY.
There is another test worth mentioning, which is the GRS test. The
GRS test in \cite{gibbons1989test} is usually an excellent procedure
for testing intercepts. However, these two tests are not appropriate
in the high-dimensional regression setting as in our case. To be specific, these
tests are implicitly based on the assumption that all companies are
only related to the same small set of basis assets. Here the ``small''
means there are many fewer basis assets than observations. Our setting
is more general since we may have more basis assets than observations,
although each company is related to only a small number of basis assets,
different companies may be related to different sets of basis assets.
The GRS is unable to handle setting.

As noted earlier, Table \ref{tab: amf_int} shows that 4.12\% of stocks
in the multi-factor model have p-values for the intercept t-test of
less than 0.05. While in the FF5 model, this percentage is 6.33\%.
After using the BHY method to control for the false discovery rate,
we can see that the q-values (the minimum false discovery rate needed
to accept that this rejection is a true discovery, see \cite{storey2002direct})
for both models are almost 1, indicating that there are no significant
non-zero intercepts. All the significance shown in the intercept tests
is likely to be false discovery. This is the evidence that both
models are consistent with the behavior of ``large-time scale''
security returns.

\begin{table}[H]
\center %
\begin{tabular}{|| *{5}{r|} |}
\hhline{|*{5}{=}}
\multirow{2}{*}{Value Range}  & \multicolumn{4}{c||}{Percentage of Stocks (\%)}\tabularnewline
\multirow{2}{*}{}  & FF5 p-val & AMF p-val & FF5 FDR q-val & AMF FDR q-val \\
\hhline{|*{5}{=|}}
0 - 0.05 & 6.33 & 4.12 & 0.00 & 0.00 \\
\hline
0.05 - 0.9 & 84.84 & 87.22 & 0.02 & 0.00 \\
\hline
0.9 - 1 & 8.83 & 8.67 & 99.98 & 100.00 \\
\hhline{|*{5}{=|}}
\end{tabular}
\caption{Intercept Test with control of false discovery rate. The first column is the value range of p-values or q-values listed in the other columns. The other 4 columns are related to p-values and False Discovery Rate (FDR) q-values for the FF5 model and the AMF model. For each column, we listed the percentage of companies with values within each value range. It is clear that nearly all rejections of zero-alpha are false discoveries. }
\label{tab: amf_int}
\end{table}

%\clearpage
\subsection{In-Sample and Out-of-Sample Goodness-of-Fit}

\begin{figure}[H]
\center \vspace{0cm}
 \includegraphics[width=0.8\textwidth]{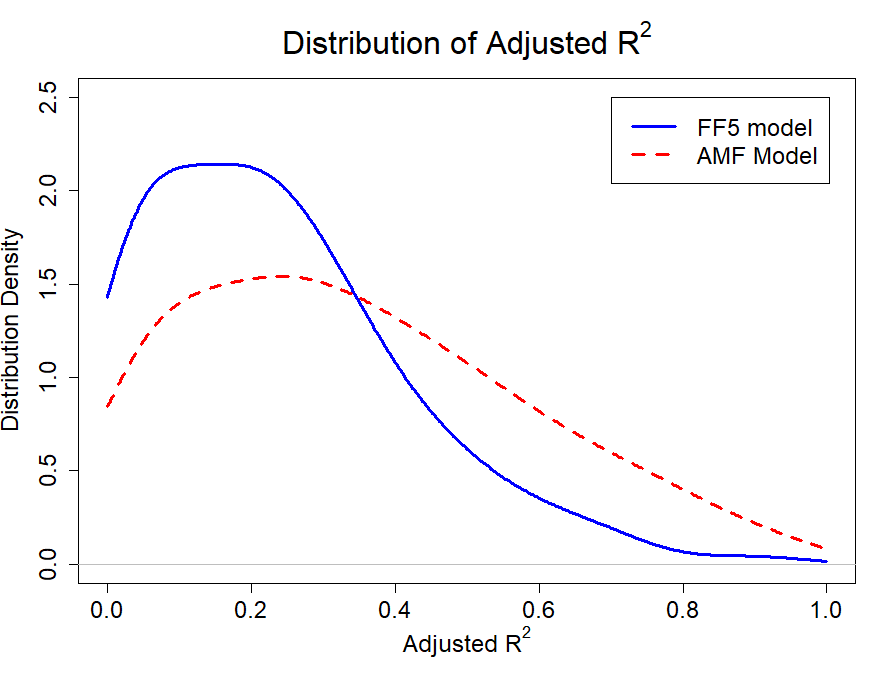}
 \captionof{figure}{Comparison
of adjusted $R^{2}$ for the Fama-French 5-factor (FF5) model and
the Adaptive Multi-factor (AMF) model}
\label{fig: amf_adj_r2}
\end{figure}

This section tests to see which model fits the data best. Figure
\ref{fig: amf_adj_r2} compares the distribution of the adjusted $R^{2}$'s (see \cite{theil1961economic}) between the AMF and the FF5 model. As indicated, the AMF model has more explanatory power. The mean adjusted $R^{2}$ for the AMF model is 0.348 while that for the FF5 model is 0.234. The AMF model increases the adjusted $R^2$ by 49\% compared to the FF5.

We next perform an F-test, for each security, to show that there is
a significant difference between the goodness-of-fit of the AMF %(equation \ref{eq: amf_ols_subset})
and the FF5 model. %(equation \ref{eq: ff5}).
Since we need a nested comparison for an F-test, we compare the results
between FF5 and GIBS + FF5 (which is including FF5 factors back to
GIBS for fitting if any of the FF5 factors are not selected). In our
case, the FF5 is the restricted model, having $p-r_{1}$ degrees of
freedom and a sum of squared residuals $SS_{R}$, where $r_{1}=5$.
The AMF is the full model, having $p-r_{1}-r_{2}$ degrees of freedom
(where $r_{2}$ is the number of basis assets selected in addition
to FF5) and a sum of squared residuals $SS_{F}$. Under the null-hypothesis
that FF5 is the true model, we have
\begin{equation}
F_{obs}=\frac{(SS_{R}-SS_{F})/r_{2}}{SS_{F}/(p-r_{1}-r_{2})}\overset{H_{0}}{\sim}F_{r_{2},p-r_{1}-r_{2}}.
\end{equation}

There are 4881 stocks in total. For 1459 (30\%) of them, the GIBS
algorithm only selects some of the FF5 factors, so for these stocks,
GIBS + FF5 does not give extra information. However, for 3422 (70\%)
of them, the GIBS algorithm does select ETFs outside of the FF5 factors.
For these stocks, we do the F test to check whether the difference
between the two models are significant, in other words, whether AMF
gives a significantly better fit. As shown in Table \ref{tab: amf_f_test},
for 99.24\% of the stocks, the AMF model fits better than the FF5
model.

Again, it is important to test the False Discovery Rate (FDR). Table
\ref{tab: amf_f_test} contains the p-values and the false discovery rate
q-values using both the BH method and BHY methods. As indicated, for
most of the stocks, the AMF is significantly better than the FF5 model,
even after considering the false discovery rate. For 99.24\% of stocks,
the AMF model is better than the FF5 at the significance level of
0.05. After considering the false discovery rate using the strict
BHY method which includes the correlation between tests, there is
still 87.93\% of stocks significant with q-values less than 0.05.
Even if we adjust our false discover rate q-value significance level
to 0.01, there is still 74.08\% of the stocks showing a significant
difference. As such, this is strong evidence supporting the multi-factor
model's superior performance in characterizing security returns.

\begin{table}[H]
\center %
\begin{tabular}{||c|r|r|r||}
\hhline{|=|===|}
\multirow{2}{*}{Value Range}  & \multicolumn{3}{c||}{Percentage of Stocks}\tabularnewline
\multirow{2}{*}{}  & p-value  & BH q-value  & BHY q-value\tabularnewline
\hline
$ 0 \sim 0.01 $ & 93.19\% & 92.78\% & 74.08\% \\
\hline
$ 0 \sim 0.05 $ (Significant) & 99.24\% & 99.24\% & 87.93\% \\
\hline
$ 0.05 \sim 1 $ (Non-Significant) & 0.76\% & 0.76\% & 12.07\% \\
\hhline{|=|=|=|=|}
\end{tabular}\caption{F test with control of false discovery rate. We do the F test and
report its p-value, q-values for each company. The first column is
the value range of p-values or q-values listed in the other columns.
In the other three columns we report percentage of companies with
p-value, BH method q-value, and BHY method q-value in each value range.
The table shows that for most companies the increment of goodness
of fit is very significant.}
\label{tab: amf_f_test}
\end{table}

Apart from the In-Sample goodness of fit results, we also compare
the Out-of-Sample goodness of fit of the FF5 and AMF model in the
prediction time period. We use the two models to predict the return
of the following week and report the Out-of-Sample $R^{2}$ for the
prediction (see Table \ref{tab: amf_compare_methods}). The Out-of-Sample
$R^{2}$ (see \cite{campbell2008predicting}) is used to measure the
predictive accuracy of a model. The Out-of-Sample $R^{2}$ for the
FF5 is $0.209$, while that for the AMF is $0.331$. That is, the
AMF model increased the Out-of-Sample $R^{2}$ by $58\%$ compared
to the FF5 model. The AMF model shows its superior performance by
giving a more accurate prediction using an even lower number of factors,
which is also strong evidence against overfitting. The AMF model provides
additional insight when compared to the FF5.

%\clearpage
\subsection{Robustness Test}

As a robustness test, for those securities whose intercepts were non-zero,
we tested the basis asset implied multi-factor model to see if positive
alpha trading strategies generate arbitrage opportunities. To construct
the positive alpha trading strategies, we use the data from the year
2017 as an out-of-sample period. Recall that the previous analysis
was over the time period 2014 to 2016. As explained above, we fit
the AMF model using the data up to the last week of 2016. We then
ranked the securities by their alphas from positive to negative. We
take the top 50\% of those with significant (p-val less than 0.05)
positive alphas and form a long-only equal-weighted portfolio with
\$1 in initial capital. Similarly, take the bottom 50\% of those with
significant negative alphas and form a short-only equal-weighted portfolio
with -\$1 initial capital. Then, each week over 2017, we update the
two portfolios by re-fitting the AMF model and repeating the same
construction. Combining the long-only and short-only portfolio forms
a portfolio with 0 initial investment. If the alphas represent arbitrage
opportunities, then the combined long and short portfolio's change
in value will always be non-negative and strictly positive for some
time periods.

\begin{figure}[H]
\center \includegraphics[width=0.65\linewidth]{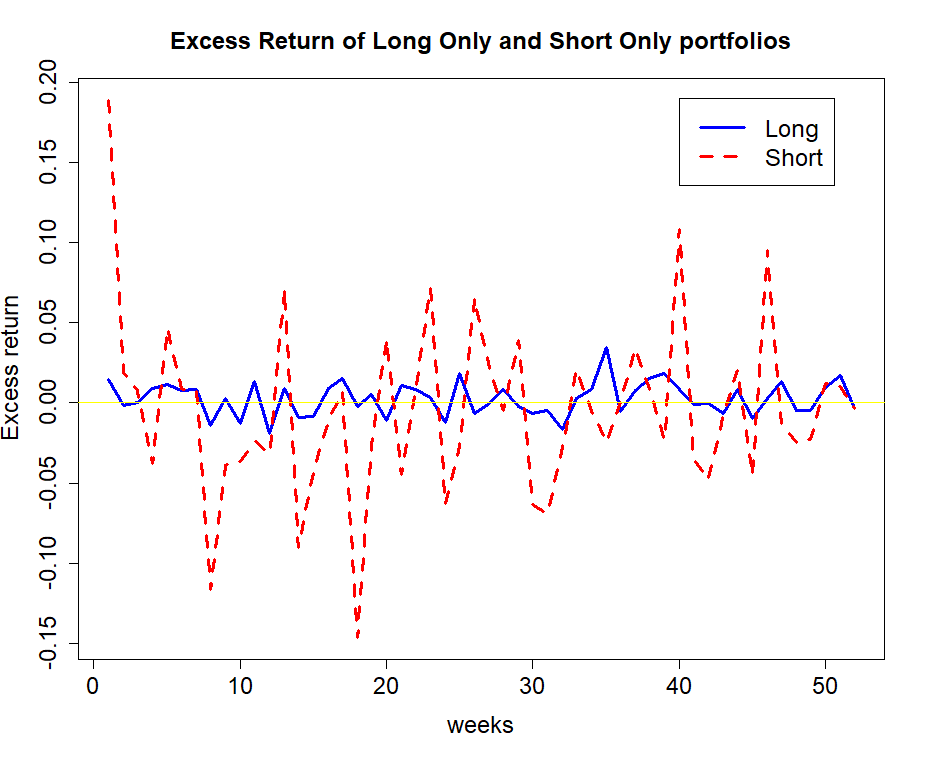} \captionof{figure}{Returns of Long-only and Short-only Portfolio}
\label{fig: amf_returns}
\end{figure}

\begin{figure}[H]
\center \includegraphics[width=0.65\linewidth]{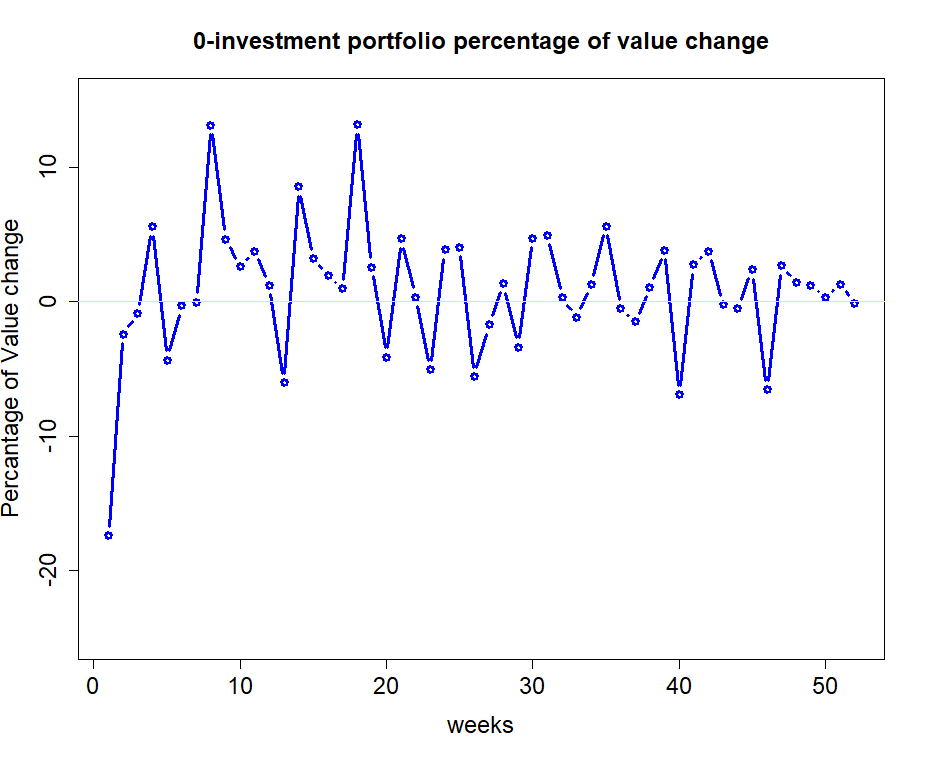}
\captionof{figure}{Percentage of Value Change of 0-Investment portfolio}
\label{fig: amf_value_change}
\end{figure}

The results of the arbitrage tests are shown in Figures \ref{fig: amf_returns}
and \ref{fig: amf_value_change}. As indicated, the change in value of the
0-investment portfolio randomly fluctuates on both sides of 0. This
rejects the possibility that the positive alpha trading strategy is
an arbitrage opportunity. Thus, this robustness test confirms our
previous intercept test results, after controlling for a false discovery
rate. Although not reported, we also studied different quantiles from
10\% to 40\% and they give similar results.

%\clearpage{}

%\begin{figure}
%\vspace{0cm}
% %
%\begin{minipage}[c]{0.5\textwidth}%
% \centering \includegraphics[width=0.99\linewidth]{adj_r2.png} \captionof{figure}{Comparison
%of Adjusted R squared} \label{fig: amf_adj_r2} %
%\end{minipage}%
%\begin{minipage}[c]{0.5\textwidth}%
% \centering \includegraphics[width=0.99\linewidth]{InterceptTest.png}
%\captionof{figure}{Comparison of intercept test} \label{fig: amf_int} %
%\end{minipage}
%\end{figure}

%\clearpage{}

\section{Comparison with Alternative Methods}
\label{sec: amf_comparison}

\subsection{Are Fama-French 5 Factors Overfitting?}

We first test whether the Fama-French 5 factors (FF5) overfit the
noise in the data. This can be done by estimating a ``GIBS + FF5''
model. This model is very similar to GIBS, except that it includes
the Fama-French 5 factors to be selected in the last step. That is,
if any of the FF5 factors are not selected by GIBS, we add them back
to our selected basis asset set $\w S_{i}$ and use this set of basis
assets to fit and predict the returns. By comparing the In-Sample
Adjusted $R^{2}$ and the Out-of-Sample $R^{2}$ (see \cite{campbell2008predicting})
of GIBS + FF5 model and the GIBS model, we can determine whether FF5
factors are overfitting. The Out-of-Sample $R^{2}$ (see \cite{campbell2008predicting})
is used to measure the accuracy of prediction of a model. Surprisingly,
the results show that some of the FF5 factors are over-fitting! As
shown in Table \ref{tab: amf_compare_methods}, compared to our GIBS model,
the GIBS + FF5 achieves a better in-sample Adjusted $R^{2}$, with
more significant basis assets, but gives a much worse Out-of-Sample
$R^{2}$. This indicates that the FF5 factors not selected by GIBS
are ``false discoveries'' - they overfit the training data, but
do a poor job in predicting. Therefore, those FF5 factors should not
be used for a company if they are not selected by GIBS. Table \ref{tab: amf_compare_methods}
not only provides evidence of the superior performance of GIBS over
FF5 by comparing the In-Sample Adjusted $R^{2}$ and Out-of-Sample
$R^{2}$ of GIBS and FF5 model, but it also indicates an overfitting
of FF5 by comparing the In-Sample Adjusted $R^{2}$ and Out-of-Sample
$R^{2}$ of GIBS and GIBS + FF5.

\subsection{Comparison with Elastic Net}

\begin{figure}[H]
\center \vspace{0cm}
\includegraphics[width=0.67\textwidth]{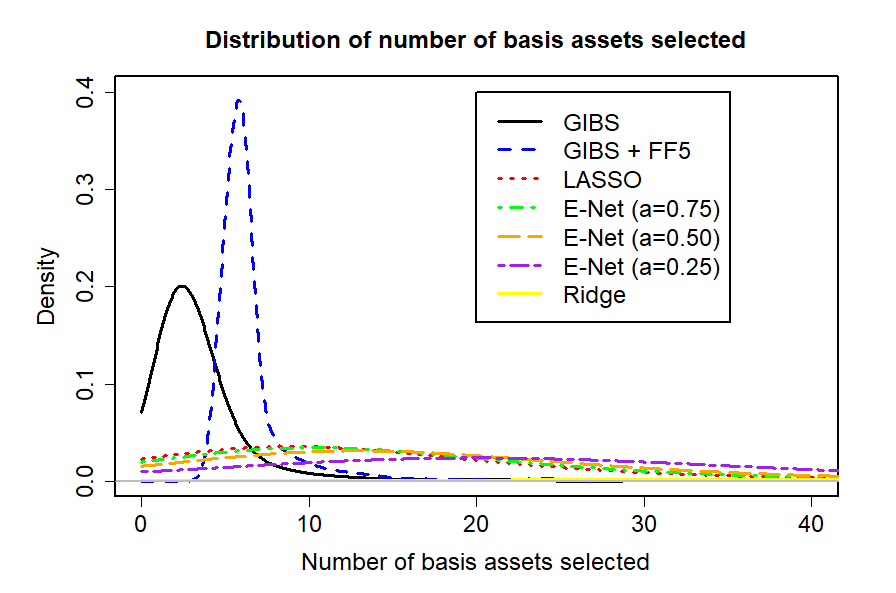}
\caption{Comparison of number of basis assets selected by cross-validation for different methods.}
\label{fig: amf_compare_n_factors}
\end{figure}

Since there are a large number of correlated ETFs it is natural to
employ the RIDGE, LASSO and Elastic Net (E-Net) methods (by Zou and
Hastie (2005) \cite{zou2005regularization}). E-Net is akin to the
ridge regression's treatment of multicollinearity with an additional
tuning (ridge) parameter, $\alpha$, that regularizes the correlations.
We compare GIBS, LASSO, RIDGE and E-Net with different $\alpha$s.
The sparsity inducing parameter $\lambda$ in each model is selected
by the usual 10-fold cross-validation (by Kohavi et al. (1995) \cite{kohavi1995study}).
The comparison results are shown in Table \ref{tab: amf_compare_methods}.
The distribution of the number of basis assets selected by each method
is shown in Figure \ref{fig: amf_compare_n_factors}.

From Table \ref{tab: amf_compare_methods} it is clear that the GIBS model
has better prediction than the FF5 model. The GIBS model increased
the Out-of-Sample $R^{2}$ by 58\% compared to FF5. Across all
models, GIBS has the highest Out-of-Sample $R^{2}$, which supports
the fact that the better In-Sample Adjusted $R^{2}$ achieved by the
other models (LASSO, RIDGE, E-Net etc.) is due to overfitting. Furthermore,
from Table \ref{tab: amf_compare_methods} and Figure \ref{fig: amf_compare_n_factors},
we see that GIBS selects the least number of factors.

\begin{table}[H]
%\resizebox{6in}{!}{
\centering
\begin{adjustbox}{width={0.95\textwidth}, totalheight={0.95\textheight},keepaspectratio}
\begin{tabular}{| *{5}{|r} ||}
\hhline{*{5}{|=}|}
Model & Select & Signif. & In-Sample Adj. $R^2$  & Out-of-Sample $R^2$ \\
\hhline{*{5}{|=}|}
FF5 & 5.0 & 1.8 & 0.234 (00\%) & 0.209 (00\%) \\
\hline
\textbf{GIBS} & \textbf{3.4} & 2.2 & 0.347 (49\%) & \textbf{0.331 (58\%)} \\
\hline
GIBS + FF5 & 6.6 & 2.5 & 0.352 (50\%) & 0.213 (02\%) \\
\hline
LASSO & 13.4 & NA & 0.372 (59\%) & 0.294 (40\%) \\
\hline
E-Net ($\alpha$=0.75) & 15.7 & NA & 0.370 (58\%) & 0.296 (41\%) \\
\hline
E-Net ($\alpha$=0.50) & 17.3 & NA & 0.364 (56\%) & 0.300 (43\%) \\
\hline
E-Net ($\alpha$=0.25) & 24.8 & NA & 0.344 (47\%) & 0.225 (08\%) \\
\hline
Ridge & 186.0 & NA & NA & 0.329 (57\%) \\
\hhline{*{5}{|=}|}
\end{tabular}
\end{adjustbox}
\caption{Comparison table for Alternative Methods. The ``Select'' column gives the average count of the factors selected by the model. The ``Signif.'' column gives the average count of the significant factors selected by the model. The column ``In-sample Adj. $R^2$'' gives the average in-sample Adjusted $R^2$ for each model, the percentage in the bracket is the percentage change compared to the FF5 model. The column ``Out-of-Sample $R^2$'' gives the average Out-of-Sample $R^2$ for each model, the percentage in the bracket is the percentage change compared to the FF5 model.}
\label{tab: amf_compare_methods}
\end{table}

The $\lambda$ of LASSO is traditionally selected by cross-validation
and with this $\lambda$, the model selects an average of 13.4 factors,
as shown in Table \ref{tab: amf_compare_methods}. However, most of the
factors selected by cross-validation are ``false-positive''. Therefore,
instead of the cross-validation, we use the ``1se rule'' with a
hard threshold 20 basis assets at most. The ``1se rule'' use the
largest $\lambda$ such that the cross-validation error is within
one standard error of the minimum error achieved by the cross-validation.
In another word, $\lambda_{1se}$ gives the most regularized model
such that error is within one standard error of the minimum error
achieved by the cross-validation (see \cite{friedman2010regularization,simon2011Regularization,tibshirani2012strong}).
To further avoid over-fitting, we include the threshold that each
company can not be related to more than 20 basis assets. As shown
in the Table \ref{tab: amf_compare_methods}, our method used in GIBS
works well and achieves the best prediction power. The reason for
the superior performance of GIBS compared with the cross-validation
LASSO is that cross-validation often overfits, especially when the
sample size is small, or when the data is not sufficiently independent
and identically distributed. In addition, our results with GIBS are
both stable and interpretable.

Our choice of dimension reduction techniques, using a combination
of prototype clustering and a modified version of LASSO, was motivated
by our desire to select, from a collection of strongly correlated
ETFs, a sparse and interpretable set of basis assets that explains
the cross-sectional variation among asset returns. These two steps
were used as model selection tools to identify basis assets, and we
subsequently estimated the model coefficients using OLS. In future
research, a more integrative method may be designed to combine the
model selection and estimation steps.

The motivation for using prototype clustering is two fold. First,
it can be used to derive the cluster structure of the ETFs so that
the redundant ones are removed. This reduces the correlation and validates
the use of LASSO. Second, this method gives a clear interpretation
of the prototypes, which is important for our interpretation of the
basis assets. The traditional methods of dealing with empirical asset
models are based on variance decomposition of the basis assets (the
$X$ matrix) as in, for example, the principal component analysis
(PCA) approach. More recently, there are modern statistical methods
that introduce sparsity and high-dimensional settings in these traditional
methods (see Zou et al. (2006) \cite{zou2006sparse}). However, as
we argued in the introduction, these methods are not optimal for basis
asset models due to their difficulties in interpreting the basis assets.
Furthermore, it is the correlation that is important in the determination
of the basis assets not the variance itself. Therefore, methods that
focus on finding the rotation with the largest variance (like PCA)
are not optimal in this setting. Instead, we use correlation as our
metric in the prototype clustering step, which gives a clearer interpretation
and a direct analysis of the candidate basis assets, rather than linear
combinations of the basis assets. For this reason we believe that
prototype clustering is preferred to PCA in this setting.

For future work, modern refinements of model-selection and inference
methods may be used. For high-dimensional models obtaining valid p-values
is difficult. This is in part due to the fact that fitting a high-dimensional
model often requires penalization and complex estimation procedures,
which implies that characterizing the distribution of such estimators
is difficult. For statistical testing in the presence of sparsity
a number of new methods are appearing in the literature. One alternative
method is the post-selection procedure by Tibshirani et al. (2016)
\cite{tibshirani2016exact}. Another approach for constructing frequentistp
p-values and confidence intervals for high-dimensional models uses
the idea of de-biasing which was proposed in a series of articles
\cite{vandegeer2014asymptotically, javanmard2014confidence, zhang2014confidence}.
In the de-biasing method, starting from a regularized estimator one
first constructs a de-biased estimator and then makes inference based
on the asymptotic normality of low-dimensional functionals of the
de-biased estimator. In principle, this approach also provides asymptotically
valid p-values for hypotheses testing each of the parameters in the
model. However, in our numerical explorations of these methods we
found that the confidence intervals are too large for the current
application and no meaningful insights could be obtained. The p-values
for these methods are also not very stable. So we use the OLS after
LASSO instead of the post-selection methods with the theoretical guarantee
in the paper by Zhao et al. (2017) \cite{zhao2017defense}.

%\clearpage
\section{Risk-Factor Determination}

\label{sec: amf_risk_premium}

We focus on the same three year time period 2014 - 2016 and compute the average annal excess returns on the basis assets to determine which have non-zero risk premium (average excess returns), i.e. which are risk-factors in the traditional sense. In this time period there are 186 basis assets selected, including the Fama-French 5 factors and 181 ETFs. The risk premium of the Fama-French 5 factors are shown in Table \ref{tab: amf_ff5_risk_premium}.

\begin{table}[H]
\centering
\begin{tabular}{||r|r|r|r|r|r||}
  \hhline{|=|=|=|=|=|=|}
Fama-French 5 Factors & Market Return & SMB & HML & RMW & CMA \\
  \hline
Annual Excess Return (\%) & 10.0 & -2.1 & 1.7 & 1.2 & -1.1 \\
   \hhline{|=|=|=|=|=|=|}
\end{tabular}
\caption{Risk Premium of Fama-French 5 factors}
\label{tab: amf_ff5_risk_premium}
\end{table}

Out of the 181 selected ETFs, 141 of them have absolute risk premiums larger than the minimum of that of the FF5 factors (which is the RMW, with absolute risk premium 1.1\%). Therefore, at least $ (141 + 5) / (181 + 5) = 78\% $ basis assets are risk factors. Furthermore, 37 out of the 181 selected ETFs have absolute risk premiums larger than 10.0\% (the absolute risk premium of the market return, which is the biggest absolute risk premium of all FF5 factors). The list of the 37 ETFs are in Table \ref{tab: amf_etf_risk_premium}.

\begin{table}[H]
%\resizebox{6in}{!}{
\centering
\begin{adjustbox}{width={0.95\textwidth}, totalheight={0.95\textheight},keepaspectratio}
\begin{tabular}{||l|l|r||}
  \hhline{|=|=|=|}
 & & Risk Pre-  \hspace{0.05cm}  \\
ETF name & Category & mium (\%) \\
\hhline{|=|=|=|}
ProShares Ultra Bloomberg Natural Gas & Leveraged Commodities & -30.4 \\
\hline
iShares PHLX Semiconductor ETF & Technology Equities & 26.4 \\
\hline
ProShares Ultra Real Estate & Leveraged Real Estate & 26.2 \\
\hline
Global X MSCI Nigeria ETF & Foreign Large Cap Equities & -24.0 \\
\hline
Global X FTSE Greece 20 ETF & Emerging Markets Equities & -20.8 \\
\hline
First Trust Indxx Global Agriculture ETF & Large Cap Blend Equities & -20.6 \\
\hline
Invesco High Yield Equity Dividend Achievers ETF & All Cap Equities & 20.4 \\
\hline
Invesco S\&P SmallCap Information Technology ETF & Technology Equities & 19.7 \\
\hline
ProShares Short VIX Short-Term Futures & Leveraged Volatility & 18.8 \\
\hline
ProShares Ultra Consumer Services & Leveraged Equities & 18.3 \\
\hline
Direxion Daily Energy Bull 3X Shares & Leveraged Equities & -17.5 \\
\hline
Invesco DB Energy Fund & Oil \& Gas & -17.3 \\
\hline
Invesco S\&P SmallCap Consumer Staples ETF & Consumer Staples Equities & 15.8 \\
\hline
SPDR S\&P Regional Banking ETF & Financials Equities & 15.5 \\
\hline
VanEck Vectors Rare Earth/Strategic Metals ETF & Materials & -15.5 \\
\hline
Global X Uranium ETF & Global Equities & -15.3 \\
\hline
NYSE Technology ETF & Technology Equities & 15.1 \\
\hline
SPDR S\&P Health Care Equipment ETF & Health \& Biotech Equities & 15.0 \\
\hline
Vanguard Utilities ETF & Utilities Equities & 15.0 \\
\hline
VanEck Vectors Egypt Index ETF & Emerging Markets Equities & -15.0 \\
\hline
iShares MSCI Colombia ETF & Latin America Equities & -14.6 \\
\hline
SPDR S\&P Insurance ETF & Financials Equities & 14.2 \\
\hline
First Trust Morningstar Dividend Leaders & Large Cap Blend Equities & 14.2 \\
\hline
iShares U.S. Aerospace \& Defense ETF & Industrials Equities & 13.8 \\
\hline
ProShares Ultra Short Basic Materials & Leveraged Equities & -13.8 \\
\hline
iShares North American Tech-Multimedia Networking ETF & Communications Equities & 13.3 \\
\hline
iShares U.S. Healthcare Providers ETF & Health \& Biotech Equities & 12.6 \\
\hline
FLAG-Forensic Accounting Long-Short ETF & Long-Short & 12.2 \\
\hline
SPDR SSGA US Large Cap Low Volatility Index ETF & Volatility Hedged Equity & 12.0 \\
\hline
Global X MSCI Portugal ETF & Europe Equities & -11.7 \\
\hline
iShares Edge MSCI USA Momentum Factor ETF & Large Cap Growth Equities & 11.5 \\
\hline
Invesco Dynamic Software ETF & Technology Equities & 11.3 \\
\hline
Vanguard Consumer Staples ETF & Consumer Staples Equities & 10.9 \\
\hline
VanEck Vectors Poland ETF & Europe Equities & -10.8 \\
\hline
VanEck Vectors Retail ETF & Consumer Discretionary Equities & 10.7 \\
\hline
VanEck Vectors Morningstar Wide Moat ETF & Large Cap Blend Equities & 10.6 \\
\hline
Invesco Russell Top 200 Equal Weight ETF & Large Cap Growth Equities & 10.2 \\
\hline
\hhline{|=|=|=|}
\end{tabular}
\end{adjustbox}
\caption{List of ETFs with large absolute risk premium.}
\label{tab: amf_etf_risk_premium}
\end{table}

%\clearpage
\section{Illustrations}
\label{sec: amf_illustration}

In this section we illustrate our multi-factor estimation process
and compare the results with the Fama-French 5-factor (FF5) model
for three securities: Adobe, Bank of America, and Intel.

\subsection{Adobe}

This section contains the results for Adobe. Using Equation (\ref{eq: ff5}),
we estimate the Fama-French 5 factor (FF5) model as shown in Table
\ref{tab: adobe_ff5}. For our Adaptive Multi-Factor (AMF) model, the final
results are shown in Table \ref{tab: adobe_amf} with the description of the
ETF basis assets selected by GIBS in Table \ref{tab: adobe_etf_sgn}. The adjusted
$R^{2}$ for FF5 is 0.39, while the adjusted $R^{2}$ for AMF is 0.55.
From the Tables it is clear that different significant basis assets
are selected and the ones selected by GIBS gives much better explanation
and prediction power. Only the market return is significant among
the FF5 factors, and it is the only FF5 factor selected by the AMF model.
Additionally, Adobe's returns are related to the Invesco Dynamic Software ETF and
the NYSE Technology ETF, indicating that Adobe is sensitive to risks in the technology software sector. For Adobe, both models do not have a significant intercept, indicating that the securities
are properly priced.

\begin{table}[H]
\centering
%\begin{adjustbox}{width={0.95\textwidth}, totalheight={0.95\textheight},keepaspectratio}
\begin{tabular}{|| *{5}{r|} |}
\hhline{*{5}{|=}|}
& $\beta$ & SE & t value & P-value \\
\hhline{*{5}{|=}|}
(Intercept) & 0.002 & 0.002 & 1.234 & 0.219 \\
\hline
Market Return & 1.048 & 0.123 & 8.495 & 0.000 \\
\hline
SMB & -0.167 & 0.191 & -0.873 & 0.384 \\
\hline
HML & -0.456 & 0.248 & -1.838 & 0.068 \\
\hline
RMW & -0.330 & 0.311 & -1.063 & 0.290 \\
\hline
CMA & -0.279 & 0.426 & -0.654 & 0.514 \\
\hline
\hhline{*{5}{|=}|}
\end{tabular}
%\end{adjustbox}
\caption{Adobe with the FF5 model. The column $\beta$ provides the coefficients in the OLS regression of Adobe on FF5 factors. The standard error (SE), t value, and P-value related to each coefficient are also provided.}
\label{tab: adobe_ff5}
\end{table}

\begin{table}[H]
%\resizebox{6in}{!}{ %
\centering
\begin{adjustbox}{width={0.95\textwidth}, totalheight={0.95\textheight},keepaspectratio}
\begin{tabular}{|| *{5}{r|} |}
\hhline{*{5}{|=}|}
& $\beta$ & SE & t value & P-value \\
\hhline{*{5}{|=}|}
(Intercept) & 0.002 & 0.002 & 0.955 & 0.341 \\
\hline
Market Return & -0.481 & 0.312 & -1.539 & 0.126 \\
\hline
iShares Edge MSCI USA Momentum Factor ETF & 0.569 & 0.253 & 2.249 & 0.026 \\
\hline
Invesco Dynamic Software ETF & 0.420 & 0.175 & 2.395 & 0.018 \\
\hline
NYSE Technology ETF & 0.559 & 0.186 & 3.007 & 0.003 \\
\hline
AGFiQ US Market Neutral Value Fund & -0.443 & 0.162 & -2.739 & 0.007 \\
\hhline{*{5}{|=}|}
\end{tabular}
\end{adjustbox}
\caption{Adobe with the AMF. The column $\beta$ provides the coefficients
in the second-step OLS regression of Adobe on basis assets selected
in the AMF. The standard error (SE), t value, and P-value related
to each coefficient are also provided.}
\label{tab: adobe_amf}
\end{table}

\begin{table}[H]
%\resizebox{6in}{!}{ %
\centering
\begin{adjustbox}{width={0.95\textwidth}, totalheight={0.95\textheight},keepaspectratio}
\begin{tabular}{||l|l|l||}
\hhline{|=|=|=|}
ETF Name  & Category  & Big Class \\
\hhline{|=|=|=|}
iShares Edge MSCI USA Momentum Factor ETF & Large Cap Growth Equities & Equity \\
\hline
Invesco Dynamic Software ETF & Technology Equities & Equity \\
\hline
NYSE Technology ETF & Technology Equities & Equity \\
\hline
AGFiQ US Market Neutral Value Fund & Long-Short & Alternative ETFs \\
\hhline{|=|=|=|}
\end{tabular}
\end{adjustbox}
\caption{Significant ETF basis assets for Adobe. This table shows the category
and big class of each ETF basis asset selected in the AMF.}
\label{tab: adobe_etf_sgn}
\end{table}

\subsection{Bank of America}

This section contains the results for the Bank of America (BOA). The
results are found in Tables \ref{tab: boa_ff5}, \ref{tab: boa_amf}, and \ref{tab: boa_etf_sgn}.
The adjusted $R^{2}$ for FF5 is 0.72 while the adjusted $R^{2}$ for AMF is 0.83. For BOA, it is related to all of the FF5 factors except SMB. It is related to the ProShares Ultra 7-10 Year Treasury, Invesco CurrencyShares Swiss Franc Trust, and First Trust Morningstar Dividend Leaders, indicating that BOA's security returns (as a bank) are subject to risks related to the term structure of interest rates. It is also related to the Direxion Daily Energy Bull 3X Shares which is correlated to the health of the economy. Finally, the $\alpha$'s in both models are not significantly different from zero, indicating that no arbitrage opportunities exist for this security.

\begin{table}[H]
\center %
%\begin{adjustbox}{width={0.95\textwidth}, totalheight={0.95\textheight},keepaspectratio}
\begin{tabular}{|| *{5}{r|} |}
\hhline{*{5}{|=}|}
& $\beta$ & SE & t value & P-value \\
\hhline{*{5}{|=}|}
(Intercept) & 0.000 & 0.002 & 0.295 & 0.769 \\
\hline
Market Return & 1.046 & 0.098 & 10.669 & 0.000 \\
\hline
SMB & 0.236 & 0.152 & 1.556 & 0.122 \\
\hline
HML & 1.980 & 0.197 & 10.033 & 0.000 \\
\hline
RMW & -1.140 & 0.247 & -4.620 & 0.000 \\
\hline
CMA & -2.057 & 0.339 & -6.077 & 0.000 \\
\hhline{*{5}{|=}|}
\end{tabular}
\caption{BOA with the FF5 model. The column $\beta$ provides the coefficients in the OLS regression of BOA on FF5 factors. The standard error (SE), t value, and P-value related to each coefficient are also provided.}
\label{tab: boa_ff5}
\end{table}

\begin{table}[H]
%\resizebox{6in}{!}{ %
\centering
\begin{adjustbox}{width={0.95\textwidth}, totalheight={0.95\textheight},keepaspectratio}
\begin{tabular}{|| *{5}{r|} |}
\hhline{*{5}{|=}|}
& $\beta$ & SE & t value & P-value \\
\hhline{*{5}{|=}|}
(Intercept) & 0.000 & 0.001 & 0.329 & 0.743 \\
\hline
Market Return & 1.892 & 0.197 & 9.583 & 0.000 \\
\hline
HML & 1.856 & 0.174 & 10.636 & 0.000 \\
\hline
RMW & -1.012 & 0.203 & -4.986 & 0.000 \\
\hline
CMA & -1.122 & 0.281 & -3.999 & 0.000 \\
\hline
ProShares Ultra 7-10 Year Treasury & -0.239 & 0.111 & -2.157 & 0.033 \\
\hline
AdvisorShares Newfleet Multi-Sector Income ETF & -1.111 & 0.757 & -1.468 & 0.144 \\
\hline
Invesco DB Precious Metals Fund & -0.085 & 0.071 & -1.199 & 0.233 \\
\hline
Invesco CurrencyShares Swiss Franc Trust & -0.170 & 0.075 & -2.286 & 0.024 \\
\hline
iShares MSCI Turkey ETF & -0.062 & 0.036 & -1.741 & 0.084 \\
\hline
Direxion Daily Energy Bull 3X Shares & -0.132 & 0.026 & -5.120 & 0.000 \\
\hline
First Trust S\&P International Dividend Aristocrats ETF & -0.125 & 0.118 & -1.054 & 0.294 \\
\hline
Invesco S\&P International Developed Low Volatility ETF & 0.037 & 0.180 & 0.204 & 0.839 \\
\hline
First Trust Morningstar Dividend Leaders & -0.447 & 0.194 & -2.301 & 0.023 \\
\hhline{*{5}{|=}|}
\end{tabular}
\end{adjustbox}
\caption{BOA with the AMF. The column $\beta$ provides the coefficients in
the second-step OLS regression of BOA on basis assets selected in
the AMF. The standard error (SE), t value, and P-value related to
each coefficient are also provided.}
\label{tab: boa_amf}
\end{table}

\begin{table}[H]
%\resizebox{6in}{!}{ %
\centering
\begin{adjustbox}{width={0.85\textwidth}, totalheight={0.95\textheight},keepaspectratio}
\begin{tabular}{||l|l|l||}
\hhline{|=|=|=|}
ETF Name  & Category  & Big Class \tabularnewline
\hhline{|=|=|=|}
ProShares Ultra 7-10 Year Treasury & Leveraged Bonds & Leveraged \\
\hline
Invesco CurrencyShares Swiss Franc Trust & Currency & Currency \\
\hline
Direxion Daily Energy Bull 3X Shares & Leveraged Equities & Leveraged \\
\hline
First Trust Morningstar Dividend Leaders & Large Cap Blend Equities & Equity \\
\hhline{|=|=|=|}
\end{tabular}
\end{adjustbox}
\caption{Significant ETF basis assets for BOA. This table shows the category
and big class of each ETF basis asset selected in the AMF.}
\label{tab: boa_etf_sgn}
\end{table}

\subsection{Intel}

This section gives the results for Intel, in Tables \ref{tab: intel_ff5}, \ref{tab: intel_amf},
and \ref{tab: intel_etf_sgn}. The adjusted $R^{2}$ for FF5 is 0.45, while the adjusted $R^{2}$ for our AMF model is 0.57. For Intel, the market return and SMB are the significant FF5 factors. It is significantly related to iShares PHLX Semiconductor ETF, since Intel produces semiconductor chips. Finally, the $\alpha$'s in both models are not significantly different from zero.

\begin{table}[H]
\center %
%\begin{adjustbox}{width={0.95\textwidth}, totalheight={0.95\textheight},keepaspectratio}
\begin{tabular}{|| *{5}{r|} |}
\hhline{*{5}{|=}|}
& $\beta$ & SE & t value & P-value \\
\hhline{*{5}{|=}|}
(Intercept) & 0.001 & 0.002 & 0.483 & 0.630 \\
\hline
Market Return & 1.289 & 0.117 & 11.007 & 0.000 \\
\hline
SMB & -0.390 & 0.181 & -2.150 & 0.033 \\
\hline
HML & -0.044 & 0.236 & -0.187 & 0.852 \\
\hline
RMW & 0.499 & 0.295 & 1.691 & 0.093 \\
\hline
CMA & 0.019 & 0.404 & 0.046 & 0.963 \\
\hhline{*{5}{|=}|}
\end{tabular}\caption{Intel with the FF5 model. The column $\beta$ provides the coefficients in the OLS regression of Intel on FF5 factors. The standard error
(SE), t value, and P-value related to each coefficient are also provided.}
\label{tab: intel_ff5}
\end{table}

\begin{table}[H]
%\resizebox{6in}{!}{ %
\centering
\begin{adjustbox}{width={0.8\textwidth}, totalheight={0.95\textheight},keepaspectratio}
\begin{tabular}{|| *{5}{r|} |}
\hhline{*{5}{|=}|}
& $\beta$ & SE & t value & P-value \\
\hhline{*{5}{|=}|}
(Intercept) & 0.000 & 0.002 & -0.094 & 0.925 \\
\hline
Market Return & 0.332 & 0.170 & 1.950 & 0.053 \\
\hline
SMB & -0.486 & 0.149 & -3.255 & 0.001 \\
\hline
iShares PHLX Semiconductor ETF & 0.673 & 0.101 & 6.674 & 0.000 \\
\hhline{*{5}{|=}|}
\end{tabular}
\end{adjustbox}
\caption{Intel with AMF. The column $\beta$ provides the coefficients in the
second-step OLS regression of Intel on basis assets selected in the
AMF. The standard error (SE), t value, and P-value related to each
coefficient are also provided.}
\label{tab: intel_amf}
\end{table}

\begin{table}[H]
%\resizebox{6in}{!}{ %
\centering
\begin{adjustbox}{width={0.8\textwidth}, totalheight={0.95\textheight},keepaspectratio}
\begin{tabular}{||l|l|l||}
\hhline{|=|=|=|}
ETF Name  & Category  & Big Class \tabularnewline
\hhline{|=|=|=|}
iShares PHLX Semiconductor ETF & Technology Equities & Equity \\
\hhline{|=|=|=|}
\end{tabular}
\end{adjustbox}
\caption{Significant ETF pbasis assets for Intel. This table shows the category
and big class of each ETF basis asset selected in the AMF.}
\label{tab: intel_etf_sgn}
\end{table}

%\clearpage
\section{Conclusion}

\label{sec: amf_conclusion}

Using a collection of basis assets, the purpose of this paper is to test the new Adaptive Multi-Factor (AMF) model implied by the generalized arbitrage pricing theory (APT) recently developed by Jarrow and Protter (2016) \cite{jarrow2016positive} and Jarrow (2016) \cite{jarrow2016bubbles}. The idea is to obtain the collection of all possible basis assets and to provide a simultaneous test, security by security, of which basis assets are significant. Since the collection of basis assets selected for investigation is large and highly correlated, we propose a new high-dimensional algorithm -- the Groupwise Interpretable Basis Selection (GIBS) algorithm to do the analysis. For comparison with the existing literature, we compare the performance of AMF (using the GIBS algorithm) with the Fama-French 5-factor model and all other alternative methods. Both the Fama-French 5-factor and the AMF model are consistent with the behavior of ``large-time scale'' security returns. In a goodness-of-fit test comparing the AMF with Fama-French 5-factor model, the AMF model has a substantially larger In-Sample adjusted $R^2$ and Out-of-Sample $R^2$. This documents the AMF model's superior performance in characterizing security returns. Last, as a robustness test, for those securities whose intercepts were non-zero (although insignificant), we tested the AMF model to see if positive alpha trading strategies generate arbitrage opportunities. They do not, thereby confirming that the multi-factor model provides a reasonable characterization of security returns.

%\bibliography{bib_liao}
%\printbibliography

%\bibliographystyle{ieeetr}
%%\bibliographystyle{chicago}
\bibliographystyle{plain}
\bibliography{bib_liao}

\begin{appendices}

\section{Summary of Generalized APT}
\label{ap: summary_apt}
This appendix provides a summary of the key results
from the generalized APT contained in Jarrow and Protter (2016) \cite{jarrow2016positive}
that are relevant to the empirical estimation and different from the
traditional approach to the estimation of factor models. Prior to
that, however, we first discuss the traditional approach to the estimation
of factor models.

The traditional approach to the estimation of factor models, based
on Merton (1973) \cite{merton1973intertemporal} and Ross (1976) \cite{ross1976arbitrage},
starts with an expected return relation between a risky asset and
a finite collection of risk-factors. Merton's relation is derived
from the first order conditions of an investor's optimization problem,
given in expectations. Ross's relation is derived from a limiting
arbitrage pricing condition satisfied by the asset's expected returns.
Given these are expected return relations, the included risk factor's
excess expected returns are all non-zero by construction. These non-zero
expected excess returns are interpreted as risk premiums earned for
systematic risk exhibited by the risk factors. Jensen's alpha is then
viewed as a wedge between the risky assets expected return and those
of the risk factors implying a mispricing in the market.

To get an empirical relation in realized returns for estimation, the
traditional approach rewrites these risk-factor expected returns as
a realized return less an error. Substitution yields an expression
relating a risky asset's realized return to the realized return's
of the risk factors plus a cumulative error. Since the risk factors
do not reflect idiosyncratic risk, the cumulative error in this relation
may not satisfy the standard assumptions needed for a regression model.
In particular, the error term may be autocorrelated and/or correlated
with other idiosyncratic risk terms not included in the estimated
equation. Because the estimated relation is derived from a relation
involving expectations using risk-factors, the $R^2$ of the realized return regression can be small.

Jarrow and Protter (2016) derive a testable multi-factor model in
a different and more general setting. With trading in a finite number
for risky assets in the context of a continuous time, continuous trading
market assuming only frictionless and competitive markets that satisfy
no-arbitrage and no dominance, i. e. the existence of an equivalent
martingale measure, they are able to derive a no-arbitrage condition
satsifed by any trading strategy's realized returns. Adding additional
structure to the economy and then taking expectations yields both
Merton's (1973) \cite{merton1973intertemporal} and Ross's (1976) \cite{ross1976arbitrage}
models as special cases.

The generalize APT uses linear algebra to prove the existence of an
algebraic basis in the risky asset's payoff space at some future time
T. Since this is is a continuous time and trading economy, this payoff
space is infinite dimensional. The algebraic basis at time T constitutes
the collection of basis assets. It is important to note that this
set of basis assets are tradable. An algebraic basis means that any
risky asset's return can be written as a linear combination of a \emph{finite}
number of the basis asset returns, and different risky assets may
have a \emph{different finite }combination of basis assets explaining
their returns. Since the space of random variables generated by the
admissible trading strategies is infinite dimensional, this algebraic
basis representation of the relevant risks is parsimonious and sparse.
Indeed, only a finite number of basis assets in an infinite dimensional
space explains any trading strategy's return process.

No arbitrage, the martingale relation, implies that the same set of
basis assets explains any particular risky asset's realized returns
at all earlier times $t\in[0,T]$ as well. This is the no arbitrage
relation satisfied by the realized return processes. Adding a non-zero
alpha to this relation (Jensen's alpha) implies a violation of the
no-arbitrage condition. This no-arbitrage condition is valid and testable
in realized return space.

In contrast to the traditional approach (discussed above), deriving
the relation in realized return space implies that the error structure
is more likely to satisfy the standard assumptions of regression models.
This follows because including basis assets with zero excess expected
returns (which are excluded in the traditional approach based on risk-factors)
will reduce correlations between the error terms across time and cross-sectionally
between the error terms and the other basis assets (independent variables).
The result is the estimated equation should have a larger $R^{2}$.
After fitting the multi-factor model in realized returns, estimating
the relevant basis assets expectations determines which are risk-factors,
i.e. which earn a risk premium for systematic risk.

\clearpage
\section{Company Classes by SIC Code}
\label{sec: amf_comp_classes}

The following list of company classes based on the first two digits of Standard Industrial Classification (SIC) code is from the website of United States Department of Labor
(\url{https://www.osha.gov/pls/imis/sic_manual.html}):

{\small
\begin{longtable}{|| m{0.18\textwidth} | m{0.72\textwidth} ||}

\endhead
\multicolumn{2}{c}{{\footnotesize{} Continued on the next page }}  \\
\endfoot
\endlastfoot

\multicolumn{2}{c}{} \\
\hhline{|=|=|}
Division A  &  Agriculture, Forestry, And Fishing\\
\hhline{|=|=|}
Major Group 01 &  Agricultural Production Crops\\
\hline
Major Group 02 &  Agriculture production livestock and animal specialties\\
\hline
Major Group 07 &  Agricultural Services\\
\hline
Major Group 08 &  Forestry\\
\hline
Major Group 09 &  Fishing, hunting, and trapping\\
\hhline{|=|=|}

\multicolumn{2}{c}{} \\
\hhline{|=|=|}
Division B &  Mining\\
\hhline{|=|=|}
Major Group 10 &  Metal Mining\\
\hline
Major Group 12 &  Coal Mining\\
\hline
Major Group 13 &  Oil And Gas Extraction\\
\hline
Major Group 14 &  Mining And Quarrying Of Nonmetallic Minerals, Except Fuels\\
\hhline{|=|=|}

\multicolumn{2}{c}{} \\
\hhline{|=|=|}
Division C &  Construction\\
\hhline{|=|=|}
Major Group 15 &  Building Construction General Contractors And Operative Builders \\
\hline
Major Group 16 &  Heavy Construction Other Than Building Construction Contractors \\
\hline
Major Group 17 &  Construction Special Trade Contractors\\
\hhline{|=|=|}

\multicolumn{2}{c}{} \\
\hhline{|=|=|}
Division D &  Manufacturing\\
\hhline{|=|=|}
Major Group 20 &  Food And Kindred Products\\
\hline
Major Group 21 &  Tobacco Products\\
\hline
Major Group 22 &  Textile Mill Products\\
\hline
Major Group 23 &  Apparel And Other Finished Products Made From Fabrics And Similar Materials\\
\hline
Major Group 24 &  Lumber And Wood Products, Except Furniture\\
\hline
Major Group 25 &  Furniture And Fixtures\\
\hline
Major Group 26 &  Paper And Allied Products\\
\hline
Major Group 27 &  Printing, Publishing, And Allied Industries\\
\hline
Major Group 28 &  Chemicals And Allied Products\\
\hline
Major Group 29 &  Petroleum Refining And Related Industries\\
\hline
Major Group 30 &  Rubber And Miscellaneous Plastics Products\\
\hline
Major Group 31 &  Leather And Leather Products\\
\hline
Major Group 32 &  Stone, Clay, Glass, And Concrete Products\\
\hline
Major Group 33 &  Primary Metal Industries\\
\hline
Major Group 34 &  Fabricated Metal Products, Except Machinery And \hspace{2cm} Transportation Equipment\\
\hline
Major Group 35 &  Industrial And Commercial Machinery And Computer \hspace{3cm} Equipment\\
\hline
Major Group 36 &  Electronic And Other Electrical Equipment And Components,
Except Computer Equipment\\
\hline
Major Group 37 &  Transportation Equipment\\
\hline
Major Group 38 &  Measuring, Analyzing, And Controlling Instruments;
Photographic, Medical And Optical Goods; Watches And Clocks\\
\hline
Major Group 39 &  Miscellaneous Manufacturing Industries\\
\hhline{|=|=|}

\multicolumn{2}{c}{} \\
\hhline{|=|=|}
Division E  &  Transportation, Communications, Electric, Gas, And Sanitary
Services\\
\hhline{|=|=|}
Major Group 40 &  Railroad Transportation\\ \hline
Major Group 41 &  Local And Suburban Transit And Interurban Highway
Passenger Transportation\\ \hline
Major Group 42 &  Motor Freight Transportation And Warehousing\\ \hline
Major Group 43 &  United States Postal Service\\ \hline
Major Group 44 &  Water Transportation\\ \hline
Major Group 45 &  Transportation By Air\\ \hline
Major Group 46 &  Pipelines, Except Natural Gas\\ \hline
Major Group 47 &  Transportation Services\\ \hline
Major Group 48 &  Communications\\ \hline
Major Group 49 &  Electric, Gas, And Sanitary Services\\
\hhline{|=|=|}

\multicolumn{2}{c}{} \\
\hhline{|=|=|}
Division F &  Wholesale Trade\\
\hhline{|=|=|}
Major Group 50 &  Wholesale Trade-durable Goods\\ \hline
Major Group 51 &  Wholesale Trade-non-durable Goods\\
\hhline{|=|=|}

\multicolumn{2}{c}{} \\
\hhline{|=|=|}
Division G &  Retail Trade\\
\hhline{|=|=|}
Major Group 52 &  Building Materials, Hardware, Garden Supply, And
Mobile Home Dealers\\ \hline
Major Group 53 &  General Merchandise Stores\\ \hline
Major Group 54 &  Food Stores\\ \hline
Major Group 55 &  Automotive Dealers And Gasoline Service Stations\\ \hline
Major Group 56 &  Apparel And Accessory Stores\\ \hline
Major Group 57 &  Home Furniture, Furnishings, And Equipment Stores\\ \hline
Major Group 58 &  Eating And Drinking Places\\ \hline
Major Group 59 &  Miscellaneous Retail\\
\hhline{|=|=|}

\multicolumn{2}{c}{} \\
\hhline{|=|=|}
Division H &  Finance, Insurance, And Real Estate\\
\hhline{|=|=|}
Major Group 60 &  Depository Institutions\\ \hline
Major Group 61 &  Non-depository Credit Institutions\\ \hline
Major Group 62 &  Security And Commodity Brokers, Dealers, Exchanges,
And Services\\ \hline
Major Group 63 &  Insurance Carriers\\ \hline
Major Group 64 &  Insurance Agents, Brokers, And Service\\ \hline
Major Group 65 &  Real Estate\\ \hline
Major Group 67 &  Holding And Other Investment Offices\\
\hhline{|=|=|}

\multicolumn{2}{c}{} \\
\hhline{|=|=|}
Division I &  Services\\
\hhline{|=|=|}
Major Group 70 &  Hotels, Rooming Houses, Camps, And Other Lodging
Places\\ \hline
Major Group 72 &  Personal Services\\ \hline
Major Group 73 &  Business Services\\ \hline
Major Group 75 &  Automotive Repair, Services, And Parking\\ \hline
Major Group 76 &  Miscellaneous Repair Services\\ \hline
Major Group 78 &  Motion Pictures\\ \hline
Major Group 79 &  Amusement And Recreation Services\\ \hline
Major Group 80 &  Health Services\\ \hline
Major Group 81 &  Legal Services\\ \hline
Major Group 82 &  Educational Services\\ \hline
Major Group 83 &  Social Services\\ \hline
Major Group 84 &  Museums, Art Galleries, And Botanical And Zoological
Gardens\\ \hline
Major Group 86 &  Membership Organizations\\ \hline
Major Group 87 &  Engineering, Accounting, Research, Management, And
Related Services\\ \hline
Major Group 88 &  Private Households\\ \hline
Major Group 89 &  Miscellaneous Services\\
\hhline{|=|=|}

\multicolumn{2}{c}{} \\
\hhline{|=|=|}
Division J &  Public Administration\\
\hhline{|=|=|}
Major Group 91 &  Executive, Legislative, And General Government, Except
Finance\\ \hline
Major Group 92 &  Justice, Public Order, And Safety\\ \hline
Major Group 93 &  Public Finance, Taxation, And Monetary Policy\\ \hline
Major Group 94 &  Administration Of Human Resource Programs\\ \hline
Major Group 95 &  Administration Of Environmental Quality And Housing Programs\\ \hline
Major Group 96 &  Administration Of Economic Programs\\ \hline
Major Group 97 &  National Security And International Affairs\\ \hline
Major Group 99 &  Nonclassifiable Establishments \\
\hhline{|=|=|}
\caption{Company classes by the Standard Industrial Classification (SIC) code}
\end{longtable}

}

\clearpage
\section{ETF Classes and Subclasses}
\label{ap: amf_etf_classes}

ETFs can be divided into 10 classes, 73 subclasses (categories) in total, based on their financial explanations. The classify criteria are found from the ETFdb database: www.etfdb.com. The classes and subclasses are listed below:

\begin{enumerate}
\item \textbf{Bond/Fixed Income}: California Munis, Corporate Bonds, Emerging
Markets Bonds, Government Bonds, High Yield Bonds, Inflation-Protected
Bonds, International Government Bonds, Money Market, Mortgage Backed
Securities, National Munis, New York Munis, Preferred Stock/Convertible
Bonds, Total Bond Market.

\item \textbf{Commodity}: Agricultural Commodities, Commodities, Metals,
Oil \& Gas, Precious Metals.

\item \textbf{Currency}: Currency.

\item \textbf{Diversified Portfolio}: Diversified Portfolio, Target Retirement
Date.

\item \textbf{Equity}: All Cap Equities, Alternative Energy Equities, Asia
Pacific Equities, Building \& Construction, China Equities, Commodity
Producers Equities, Communications Equities, Consumer Discretionary
Equities, Consumer Staples Equities, Emerging Markets Equities, Energy
Equities, Europe Equities, Financial Equities, Foreign Large Cap Equities,
Foreign Small \& Mid Cap Equities, Global Equities, Health \& Biotech
Equities, Industrials Equities, Japan Equities, Large Cap Blend Equities,
Large Cap Growth Equities, Large Cap Value Equities, Latin America
Equities, MLPs (Master Limited Partnerships), Materials, Mid Cap Blend
Equities, Mid Cap Growth Equities, Mid Cap Value Equities, Small Cap
Blend Equities, Small Cap Growth Equities, Small Cap Value Equities,
Technology Equities, Transportation Equities, Utilities Equities,
Volatility Hedged Equity, Water Equities.

\item \textbf{Alternative ETFs}: Hedge Fund, Long-Short.

\item \textbf{Inverse}: Inverse Bonds, Inverse Commodities, Inverse Equities,
Inverse Volatility.

\item \textbf{Leveraged}: Leveraged Bonds, Leveraged Commodities, \\
 Leveraged Currency, Leveraged Equities, Leveraged Multi-Asset, Leveraged
Real Estate, Leveraged Volati-lity.

\item \textbf{Real Estate}: Global Real Estate, Real Estate.

\item \textbf{Volatility}: Volatility.
\end{enumerate}

\clearpage
\section{Low-Correlated ETF Name List}
\label{ap: etf_representatives}

The low-correlated ETF name list in Section \ref{sec: gibs_algo} is in Table \ref{tab: amf_etf_representatives}.

{
\small
%\fontsize{1}{1.2}\selectfont

\begin{longtable}{|| m{0.61\textwidth} | m{0.31\textwidth} ||}
\hhline{|=|=|}
ETF Names  & Category  \\
\hhline{|=|=|}
\endhead
\hline
\multicolumn{2}{c}{{\footnotesize{}Continued on next page}}  \\
\endfoot
\endlastfoot
iShares California Muni Bond ETF & California Munis \\
\hline
iShares Floating Rate Bond ETF & Corporate Bonds \\
\hline
WisdomTree Emerging Markets Corporate Bond Fund & Corporate Bonds \\
\hline
SPDR Barclays Capital Investment Grade Floating Rate ETF & Corporate Bonds \\
\hline
FlexShares Ready Access Variable Income Fund & Corporate Bonds \\
\hline
Invesco International Corporate Bond ETF & Corporate Bonds \\
\hline
VanEck Vectors Investment Grade Floating Rate ETF & Corporate Bonds \\
\hline
iShares iBonds Mar 2020 Corporate ex-Financials ETF & Corporate Bonds \\
\hline
iShares iBonds Mar 2020 Corporate ETF & Corporate Bonds \\
\hline
iShares Emerging Markets Corporate Bond ETF & Corporate Bonds \\
\hline
ProShares Investment Grade-Interest Rate Hedged & Corporate Bonds \\
\hline
Vanguard Emerging Markets Government Bond ETF & Emerging Markets Bonds \\
\hline
ProShares Short Term USD Emerging Markets Bond ETF & Emerging Markets Bonds \\
\hline
SPDR Barclays 1-3 Month T-Bill ETF & Government Bonds \\
\hline
iShares Short Treasury Bond ETF & Government Bonds \\
\hline
SPDR Portfolio Short Term Treasury ETF & Government Bonds \\
\hline
VanEck Vectors International High Yield Bond ETF & High Yield Bonds \\
\hline
Invesco Senior Loan ETF & High Yield Bonds \\
\hline
WisdomTree Interest Rate Hedged High Yield Bond Fund & High Yield Bonds \\
\hline
SPDR Blackstone/ GSO Senior Loan ETF & High Yield Bonds \\
\hline
SPDR BofA Merrill Lynch Crossover Corporate Bond ETF & High Yield Bonds \\
\hline
First Trust Senior Loan Exchange-Traded Fund & High Yield Bonds \\
\hline
WisdomTree Negative Duration High Yield Bond Fund & High Yield Bonds \\
\hline
Invesco Global Short Term High Yield Bond ETF & High Yield Bonds \\
\hline
PIMCO 0-5 Year High Yield Corporate Bond Index Fund & High Yield Bonds \\
\hline
ProShares Inflation Expectations ETF & Inflation-Protected Bonds \\
\hline
WisdomTree Asia Local Debt Fund & International Government Bonds \\
\hline
iShares Ultra Short-Term Bond ETF & Money Market \\
\hline
VanEck Vectors AMT-Free Short Municipal Index ETF & National Munis \\
\hline
VanEck Vectors AMT-Free Intermediate Municipal Index ETF & National Munis \\
\hline
Invesco VRDO Tax-Free Weekly ETF & National Munis \\
\hline
iShares S\&P Short Term National AMT-Free Bond ETF & National Munis \\
\hline
SPDR Barclays Short Term Municipal Bond & National Munis \\
\hline
VanEck Vectors Pre-Refunded Municipal Index ETF & National Munis \\
\hline
Pimco Short Term Municipal Bond Fund & National Munis \\
\hline
SPDR Barclays Capital Convertible Bond ETF & Preferred Stock/Convertible Bonds \\
\hline
iShares U.S. Preferred Stock ETF & Preferred Stock/Convertible Bonds \\
\hline
iShares International Preferred Stock ETF & Preferred Stock/Convertible Bonds \\
\hline
Invesco CEF Income Composite ETF & Total Bond Market \\
\hline
PIMCO Enhanced Short Maturity Strategy Fund & Total Bond Market \\
\hline
Franklin Short Duration U.S. Government ETF & Total Bond Market \\
\hline
iShares Short Maturity Bond ETF & Total Bond Market \\
\hline
SPDR SSgA Ultra Short Term Bond ETF & Total Bond Market \\
\hline
Invesco Chinese Yuan Dim Sum Bond ETF & Total Bond Market \\
\hline
WisdomTree Barclays Interest Rate Hedged U.S. Aggregate Bond Fund & Total Bond Market \\
\hline
WisdomTree Barclays Negative Duration U.S. Aggregate Bond Fund & Total Bond Market \\
\hline
AdvisorShares Newfleet Multi-Sector Income ETF & Total Bond Market \\
\hline
Invesco DB Agriculture Fund & Agricultural Commodities \\
\hline
Invesco DB Base Metals Fund & Metals \\
\hline
Invesco DB Energy Fund & Oil \& Gas \\
\hline
Invesco DB Precious Metals Fund & Precious Metals \\
\hline
Aberdeen Standard Physical Palladium Shares ETF & Precious Metals \\
\hline
Invesco CurrencyShares Swiss Franc Trust & Currency \\
\hline
Invesco CurrencyShares Canadian Dollar Trust & Currency \\
\hline
Invesco DB G10 Currency Harvest Fund & Currency \\
\hline
WisdomTree Brazilian Real Fund & Currency \\
\hline
First Trust Dorsey Wright People's Portfolio ETF & Diversified Portfolio \\
\hline
First Trust Multi-Asset Diversified Income Index Fund & Diversified Portfolio \\
\hline
VanEck Vectors Israel ETF & All Cap Equities \\
\hline
Invesco High Yield Equity Dividend Achievers ETF & All Cap Equities \\
\hline
Invesco Dynamic Media ETF & All Cap Equities \\
\hline
Invesco Dynamic Leisure and Entertainment ETF & All Cap Equities \\
\hline
Invesco Global Clean Energy ETF & Alternative Energy Equities \\
\hline
First Trust ISE Global Wind Energy Index Fund & Alternative Energy Equities \\
\hline
First Trust NASDAQ Clean Edge Smart Grid Infrastructure Index Fund & Alternative Energy Equities \\
\hline
Invesco WilderHill Progressive Energy ETF & Alternative Energy Equities \\
\hline
Invesco Cleantech ETF & Alternative Energy Equities \\
\hline
Invesco BLDRS Asia 50 ADR Index Fund & Asia Pacific Equities \\
\hline
VanEck Vectors Vietnam ETF & Asia Pacific Equities \\
\hline
iShares MSCI China Small-Cap ETF & Asia Pacific Equities \\
\hline
iShares MSCI Philippines ETF & Asia Pacific Equities \\
\hline
First Trust India NIFTY 50 Equal Weight ETF & Asia Pacific Equities \\
\hline
First Trust ISE Chindia Index Fund & Asia Pacific Equities \\
\hline
iShares MSCI Thailand ETF & Asia Pacific Equities \\
\hline
WisdomTree Australia Dividend Fund & Asia Pacific Equities \\
\hline
iShares MSCI New Zealand ETF & Asia Pacific Equities \\
\hline
WisdomTree India Earnings Fund & Asia Pacific Equities \\
\hline
iShares U.S. Home Construction ETF & Building \& Construction \\
\hline
Invesco Dynamic Building \& Construction ETF & Building \& Construction \\
\hline
First Trust ISE Global Engineering and Construction ETF & Building \& Construction \\
\hline
VanEck Vectors ChinaAMC CSI 300 ETF & China Equities \\
\hline
KraneShares CSI China Five Year Plan ETF & China Equities \\
\hline
Invesco Global Agriculture ETF & Commodity Producers Equities \\
\hline
SPDR S\&P Global Natural Resources ETF & Commodity Producers Equities \\
\hline
iShares North American Tech-Multimedia Networking ETF & Communications Equities \\
\hline
iShares U.S. Telecommunications ETF & Communications Equities \\
\hline
Invesco DWA Consumer Cyclicals Momentum ETF & Consumer Discretionary Equities \\
\hline
SPDR S\&P Retail ETF & Consumer Discretionary Equities \\
\hline
VanEck Vectors Gaming ETF & Consumer Discretionary Equities \\
\hline
VanEck Vectors Retail ETF & Consumer Discretionary Equities \\
\hline
First Trust NASDAQ Global Auto Index Fund & Consumer Discretionary Equities \\
\hline
IQ Global Agribusiness Small Cap ETF & Consumer Staples Equities \\
\hline
Invesco S\&P SmallCap Consumer Staples ETF & Consumer Staples Equities \\
\hline
Vanguard Consumer Staples ETF & Consumer Staples Equities \\
\hline
WisdomTree Middle East Dividend Fund & Emerging Markets Equities \\
\hline
iShares MSCI Frontier 100 ETF & Emerging Markets Equities \\
\hline
Global X FTSE Greece 20 ETF & Emerging Markets Equities \\
\hline
VanEck Vectors Russia ETF & Emerging Markets Equities \\
\hline
iShares MSCI Turkey ETF & Emerging Markets Equities \\
\hline
VanEck Vectors Egypt Index ETF & Emerging Markets Equities \\
\hline
VanEck Vectors Poland ETF & Europe Equities \\
\hline
WisdomTree Europe Hedged Equity Fund & Europe Equities \\
\hline
First Trust United Kingdom AlphaDEX Fund & Europe Equities \\
\hline
Xtrackers MSCI United Kingdom Hedged Equity Fund & Europe Equities \\
\hline
First Trust Germany AlphaDEX Fund & Europe Equities \\
\hline
iShares MSCI Ireland ETF & Europe Equities \\
\hline
Global X MSCI Portugal ETF & Europe Equities \\
\hline
Invesco KBW High Dividend Yield Financial ETF & Financials Equities \\
\hline
ProShares Global Listed Private Equity ETF & Financials Equities \\
\hline
SPDR S\&P Insurance ETF & Financials Equities \\
\hline
Invesco Global Listed Private Equity ETF & Financials Equities \\
\hline
SPDR S\&P Regional Banking ETF & Financials Equities \\
\hline
Invesco DWA Financial Momentum ETF & Financials Equities \\
\hline
VanEck Vectors Africa Index ETF & Foreign Large Cap Equities \\
\hline
iShares MSCI EAFE ETF & Foreign Large Cap Equities \\
\hline
First Trust S\&P International Dividend Aristocrats ETF & Foreign Large Cap Equities \\
\hline
Global X MSCI Nigeria ETF & Foreign Large Cap Equities \\
\hline
Invesco S\&P International Developed Momentum ETF & Foreign Large Cap Equities \\
\hline
Global X MSCI Argentina ETF & Global Equities \\
\hline
Global X Uranium ETF & Global Equities \\
\hline
iShares MSCI Peru ETF & Global Equities \\
\hline
AdvisorShares Dorsey Wright ADR ETF & Global Equities \\
\hline
ROBO Global Robotics and Automation Index ETF & Global Equities \\
\hline
iShares U.S. Pharmaceuticals ETF & Health \& Biotech Equities \\
\hline
iShares U.S. Healthcare Providers ETF & Health \& Biotech Equities \\
\hline
SPDR S\&P Health Care Equipment ETF & Health \& Biotech Equities \\
\hline
VanEck Vectors Environmental Services ETF & Industrials Equities \\
\hline
iShares U.S. Aerospace \& Defense ETF & Industrials Equities \\
\hline
iShares MSCI Japan ETF & Japan Equities \\
\hline
First Trust Morningstar Dividend Leaders & Large Cap Blend Equities \\
\hline
Invesco S\&P 500 BuyWrite ETF & Large Cap Blend Equities \\
\hline
VanEck Vectors Morningstar Wide Moat ETF & Large Cap Blend Equities \\
\hline
iShares MSCI Israel ETF & Large Cap Blend Equities \\
\hline
First Trust Indxx Global Agriculture ETF & Large Cap Blend Equities \\
\hline
AlphaClone Alternative Alpha ETF & Large Cap Growth Equities \\
\hline
Invesco NASDAQ Internet ETF & Large Cap Growth Equities \\
\hline
Global X NASDAQ 100 Covered Call ETF & Large Cap Growth Equities \\
\hline
Invesco Russell Top 200 Equal Weight ETF & Large Cap Growth Equities \\
\hline
iShares Edge MSCI USA Momentum Factor ETF & Large Cap Growth Equities \\
\hline
iShares MSCI Colombia ETF & Latin America Equities \\
\hline
VanEck Vectors Rare Earth/Strategic Metals ETF & Materials \\
\hline
iShares Global Timber \& Forestry ETF & Materials \\
\hline
Global X Lithium ETF & Mid Cap Blend Equities \\
\hline
Invesco Global Water ETF & Mid Cap Growth Equities \\
\hline
Vanguard Russell 2000 Value ETF & Small Cap Blend Equities \\
\hline
Invesco DWA NASDAQ Momentum ETF & Small Cap Growth Equities \\
\hline
SPDR S\&P Semiconductor ETF & Technology Equities \\
\hline
Invesco Dynamic Software ETF & Technology Equities \\
\hline
iShares PHLX Semiconductor ETF & Technology Equities \\
\hline
Invesco S\&P SmallCap Information Technology ETF & Technology Equities \\
\hline
NYSE Technology ETF & Technology Equities \\
\hline
iShares Transportation Average ETF & Transportation Equities \\
\hline
Vanguard Utilities ETF & Utilities Equities \\
\hline
Invesco S\&P International Developed Low Volatility ETF & Volatility Hedged Equity \\
\hline
SPDR SSGA US Large Cap Low Volatility Index ETF & Volatility Hedged Equity \\
\hline
Invesco S\&P 500 Downside Hedged ETF & Volatility Hedged Equity \\
\hline
Invesco Water Resources ETF & Water Equities \\
\hline
First Trust ISE Water Index Fund & Water Equities \\
\hline
IQ Merger Arbitrage ETF & Hedge Fund \\
\hline
WisdomTree Managed Futures Strategy Fund & Hedge Fund \\
\hline
IQ Real Return ETF & Hedge Fund \\
\hline
First Trust Morningstar Managed Futures Strategy Fund & Hedge Fund \\
\hline
Proshares Merger ETF & Hedge Fund \\
\hline
IQ Hedge Multi-Strategy Tracker ETF & Long-Short \\
\hline
FLAG-Forensic Accounting Long-Short ETF & Long-Short \\
\hline
AGFiQ US Market Neutral Value Fund & Long-Short \\
\hline
AGFiQ US Market Neutral Size Fund & Long-Short \\
\hline
AGFiQ US Market Neutral Anti-Beta Fund & Long-Short \\
\hline
AGFiQ US Market Neutral Momentum Fund & Long-Short \\
\hline
AdvisorShares Ranger Equity Bear ETF & Inverse Equities \\
\hline
Short MSCI Emerging Markets ProShares & Inverse Equities \\
\hline
ProShares Ultra 7-10 Year Treasury & Leveraged Bonds \\
\hline
ProShares Ultra High Yield & Leveraged Bonds \\
\hline
ProShares Ultra Bloomberg Natural Gas & Leveraged Commodities \\
\hline
ProShares Ultra Yen & Leveraged Currency \\
\hline
ProShares Ultra Consumer Services & Leveraged Equities \\
\hline
ProShares Ultra Short Basic Materials & Leveraged Equities \\
\hline
Direxion Daily Energy Bull 3X Shares & Leveraged Equities \\
\hline
ProShares Ultra Real Estate & Leveraged Real Estate \\
\hline
ProShares Short VIX Short-Term Futures & Leveraged Volatility \\
\hline
iShares Europe Developed Real Estate ETF & Real Estate \\
\hhline{|=|=|}
\caption{Low-correlated ETF name list in Section \ref{sec: gibs_algo}.}
\label{tab: amf_etf_representatives}
\end{longtable}
}

\end{appendices}

\end{document}